\documentclass[12pt]{article}
\usepackage{amsfonts}
\usepackage{amssymb}
\usepackage{amsmath}

\allowdisplaybreaks
\begin{document}

\author{C. Bizdadea\thanks{%
E-mail address: bizdadea@central.ucv.ro} $^{,1}$, E.\,M. Cioroianu\thanks{%
E-mail address: manache@central.ucv.ro}$\,\,^{,1}$, S.\,O. Saliu\thanks{%
E-mail address: osaliu@central.ucv.ro}$\,\,^{,1}$\\
$^{1}$Faculty of Physics, University of Craiova,\\
13 Al. I. Cuza Str., Craiova 200585, Romania\\
E.\,M. B\u{a}b\u{a}l\^{\i}c\thanks{%
E-mail address: mbabalic@central.ucv.ro}$\,\,^{,1,2}$\\
$^{2}$Department of Theoretical Physics,\\
Horia Hulubei National Institute\\
of Physics and Nuclear Engineering,\\
PO Box MG-6, Bucharest, Magurele 077125, Romania}
\title{Dual linearized gravity in $D=6$ coupled to a purely spin-two field
of mixed symmetry $(2,2)$}
\maketitle

\begin{abstract}
Under the hypotheses of analyticity, locality, Lorentz covariance, and
Poincar\'{e} invariance of the deformations, combined with the requirement
that the interaction vertices contain at most two spatiotemporal derivatives
of the fields, we investigate the consistent interactions between a single
massless tensor field with the mixed symmetry $(3,1)$ and one massless
tensor field with the mixed symmetry $(2,2)$. The computations are done with
the help of the deformation theory based on a cohomological approach, in the
context of the antifield-BRST formalism. Our result is that dual linearized
gravity in $D=6$ gets coupled to a purely spin-two field with the mixed
symmetry of the Riemann tensor such that both the gauge transformations and
first-order reducibility relations in the $\left( 3,1\right) $ sector are
changed, but not the gauge algebra.

\end{abstract}

\section{Introduction}

Tensor fields in \textquotedblleft exotic\textquotedblright\ representations
of the Lorentz group, characterized by a mixed Young symmetry type~\cite%
{curt,curt1,aul,labast,labast1,burd,zinov1}, held the attention lately on
some important issues, like the dual formulation of field theories of spin
two or higher~\cite%
{dualsp1,dualsp2,dualsp2a,dualsp2b,dualsp3,dualsp4,dualsp5}, the
impossibility of consistent cross-interactions in the dual formulation of
linearized gravity~\cite{lingr}, or a Lagrangian first-order approach~\cite%
{zinov2,zinov3} to some classes of massless or partially massive mixed
symmetry-type tensor gauge fields, suggestively resembling to the tetrad
formalism of General Relativity. An important matter related to mixed
symmetry-type tensor fields is the study of their consistent interactions,
among themselves as well as with higher-spin gauge theories~\cite%
{high1,high2,high3,high4}. The most efficient approach to this problem is
the cohomological one, based on the deformation of the solution to the
master equation~\cite{def}. Until recently, it was commonly believed that
mixed symmetry type tensor fields are rather rigid under the introduction of
consistent interactions. Nevertheless, it has been proved that some classes
of massless tensor fields with the mixed symmetry $\left( k,1\right) $,
which are particularly important in view of their duality to linearized
gravity in $D=k+3$, allow nontrivial couplings: to a vector field for $k=3$~%
\cite{prd06}, to an arbitrary $p$-form again for $k=3$~\cite{jpa08}, and to
a topological BF model for $k=2$~\cite{epjc09}.

The purpose of this paper is
to investigate the consistent interactions between a single massless tensor
field with the mixed symmetry $(3,1)$ and one massless tensor field with the
mixed symmetry $(2,2)$. Our analysis relies on the deformation of the
solution to the master equation by means of cohomological techniques with
the help of the local BRST cohomology, whose component in the $(3,1)$ sector
has been reported in detail in~\cite{noijhep31} and in the $(2,2)$ sector
has been investigated in~\cite{r22,r22th}. Apart from the duality of the massless tensor field
with the mixed symmetry $(3,1)$ to the Pauli--Fierz theory (linearized limit
of Einstein--Hilbert gravity) in $D=6$ dimensions, it is interesting to
mention the developments concerning the dual formulations of linearized
gravity from the perspective of $M$-theory~\cite{mth1,mth2,mth3}. On the
other hand, the massless tensor field with the mixed symmetry $(2,2)$
displays all the algebraic properties of the Riemann tensor, describes
purely spin-two particles, and also provides a dual formulation of linearized
gravity in $D=5$. Actually, there is a revived interest in the construction of
dual gravity theories, which led to several new results, viz. a dual formulation of
linearized gravity in first order tetrad formalism in arbitrary dimensions
within the path integral framework~\cite{sivakumar} or a reformulation of
non-linear Einstein gravity in terms of the dual graviton together with the
ordinary metric and a shift gauge field~\cite{boulangerhohm}.

Under the hypotheses of analyticity, locality, Lorentz covariance, and
Poincar\'{e} invariance of the deformations, combined with the requirement
that the interaction vertices contain at most two spatiotemporal derivatives
of the fields, we prove that there exists a case where the deformation of
the solution to the master equation provides non-trivial cross-couplings.
This case corresponds to a six-dimensional spacetime and is described by a
deformed solution that stops at order two in the coupling constant. In this
way we establish a new result, namely that dual linearized gravity in $D=6$
gets coupled to a purely spin-two field with the mixed symmetry of the
Riemann tensor. The interacting Lagrangian action contains only
mixing-component terms of order one and two in the coupling constant. This
is the first time when both the gauge transformations and first-order
reducibility functions of the tensor field $(3,1)$ are modified at order one
in the coupling constant with terms characteristic to the $(2,2)$ sector. On
the contrary, the tensor field with the mixed symmetry $(2,2)$ remains rigid
at the level of both gauge transformations and reducibility functions. The
gauge algebra and the reducibility structure of order two are not modified
during the deformation procedure, being the same like in the case of the
starting free action. It is interesting to note that if we require the PT
invariance of the deformed theory, then no interactions occur.

\section{Free model. BRST symmetry}

We begin with the free Lagrangian action%
\begin{equation}
S_{0}\left[ t_{\lambda \mu \nu |\alpha },r_{\mu \nu |\alpha \beta }\right]
=S_{0}\left[ t_{\lambda \mu \nu |\alpha }\right] +S_{0}\left[ r_{\mu \nu
|\alpha \beta }\right] ,  \label{rt1b}
\end{equation}%
in $D\geq 5$ spatiotemporal dimensions, where%
\begin{eqnarray}
S_{0}\left[ t_{\lambda \mu \nu |\alpha }\right] &=&\int d^{D}x\left\{ \frac{1%
}{2}\left[ \left( \partial ^{\rho }t^{\lambda \mu \nu |\alpha }\right)
\left( \partial _{\rho }t_{\lambda \mu \nu |\alpha }\right) -\left( \partial
_{\alpha }t^{\lambda \mu \nu |\alpha }\right) \left( \partial ^{\beta
}t_{\lambda \mu \nu |\beta }\right) \right] \right.  \notag \\
&&-\frac{3}{2}\left[ \left( \partial _{\lambda }t^{\lambda \mu \nu |\alpha
}\right) \left( \partial ^{\rho }t_{\rho \mu \nu |\alpha }\right) +\left(
\partial ^{\rho }t^{\lambda \mu }\right) \left( \partial _{\rho }t_{\lambda
\mu }\right) \right]  \notag \\
&&\left. +3\left( \partial _{\alpha }t^{\lambda \mu \nu |\alpha }\right)
\left( \partial _{\lambda }t_{\mu \nu }\right) +3\left( \partial _{\rho
}t^{\rho \mu }\right) \left( \partial ^{\lambda }t_{\lambda \mu }\right)
\right\}  \label{rt1a}
\end{eqnarray}%
and%
\begin{eqnarray}
S_{0}\left[ r_{\mu \nu |\alpha \beta }\right] &=&\int d^{D}x\left[ \frac{1}{8%
}\left( \partial ^{\lambda }r^{\mu \nu |\alpha \beta }\right) \left(
\partial _{\lambda }r_{\mu \nu |\alpha \beta }\right) -\frac{1}{2}\left(
\partial _{\mu }r^{\mu \nu |\alpha \beta }\right) \left( \partial ^{\lambda
}r_{\lambda \nu |\alpha \beta }\right) \right.  \notag \\
&&-\left( \partial _{\mu }r^{\mu \nu |\alpha \beta }\right) \left( \partial
_{\beta }r_{\nu \alpha }\right) -\frac{1}{2}\left( \partial ^{\lambda
}r^{\nu \beta }\right) \left( \partial _{\lambda }r_{\nu \beta }\right)
\notag \\
&&\left. +\left( \partial _{\nu }r^{\nu \beta }\right) \left( \partial
^{\lambda }r_{\lambda \beta }\right) -\frac{1}{2}\left( \partial _{\nu
}r^{\nu \beta }\right) \left( \partial _{\beta }r\right) +\frac{1}{8}\left(
\partial ^{\lambda }r\right) \left( \partial _{\lambda }r\right) \right] .
\label{rt2}
\end{eqnarray}%
Everywhere in this paper we employ the flat Minkowski metric of `mostly
plus' signature $\sigma ^{\mu \nu }=\sigma _{\mu \nu }=(-++++\cdots )$. The
massless tensor field $t_{\lambda \mu \nu |\alpha }$ has the mixed symmetry $%
\left( 3,1\right) $, and hence transforms according to an irreducible
representation of $GL\left( D,\mathbb{R}\right) $ corresponding to a 4-cell
Young diagram with two columns and three rows. It is thus completely
antisymmetric in its first three indices and satisfies the identity $t_{%
\left[ \lambda \mu \nu |\alpha \right] }\equiv 0$. Here and in the sequel
the notation $[\lambda \cdots \alpha ]$ signifies complete antisymmetry with
respect to the (Lorentz) indices between brackets, with the conventions that
the minimum number of terms is always used and the result is never divided
by the number of terms. The trace of $t_{\lambda \mu \nu |\alpha }$ is
defined by $t_{\lambda \mu }=\sigma ^{\nu \alpha }t_{\lambda \mu \nu |\alpha
}$ and it is obviously an antisymmetric tensor. The massless tensor field $%
r_{\mu \nu |\alpha \beta }$ of degree four has the mixed symmetry of the
linearized Riemann tensor, and hence transforms according to an irreducible
representation of $GL\left( D,\mathbb{R}\right) $, corresponding to the
rectangular Young diagram $\left( 2,2\right) $ with two columns and two
rows. Thus, it is separately antisymmetric in the pairs $\left\{ \mu ,\nu
\right\} $ and $\left\{ \alpha ,\beta \right\} $, is symmetric under the
interchange of these pairs ($\left\{ \mu ,\nu \right\} \longleftrightarrow
\left\{ \alpha ,\beta \right\} $), and satisfies the identity $r_{\left[ \mu
\nu |\alpha \right] \beta }\equiv 0$ associated with the above diagram. The
notation $r_{\nu \beta }$ signifies the trace of the original tensor field, $%
r_{\nu \beta }=\sigma ^{\mu \alpha }r_{\mu \nu |\alpha \beta }$, which is
symmetric, $r_{\nu \beta }=r_{\beta \nu }$, while $r$ denotes its double
trace, $r=\sigma ^{\nu \beta }r_{\nu \beta }\equiv r_{\quad |\mu \nu }^{\mu
\nu }$,\ which is a scalar.

A generating set of gauge transformations for action (\ref{rt1b}) can be
taken of the form
\begin{eqnarray}
\delta _{\epsilon ,\chi }t_{\lambda \mu \nu |\alpha } &=&3\partial _{\alpha
}\epsilon _{\lambda \mu \nu }+\partial _{\left[ \lambda \right. }\epsilon
_{\left. \mu \nu \right] \alpha }+\partial _{\left[ \lambda \right. }\chi
_{\left. \mu \nu \right] |\alpha },  \label{tv7a} \\
\delta _{\xi }r_{\mu \nu |\alpha \beta } &=&\partial _{\mu }\xi _{\alpha
\beta |\nu }-\partial _{\nu }\xi _{\alpha \beta |\mu }+\partial _{\alpha
}\xi _{\mu \nu |\beta }-\partial _{\beta }\xi _{\mu \nu |\alpha }.
\label{tv7b}
\end{eqnarray}%
The gauge parameters $\epsilon _{\lambda \mu \nu }$ determine a completely
antisymmetric tensor, while the gauge parameters $\chi _{\mu \nu |\alpha }$
and $\xi _{\mu \nu |\alpha }$ display the mixed symmetry $\left( 2,1\right) $%
, such that they are antisymmetric in the first two indices and satisfy the
identities $\chi _{\left[ \mu \nu |\alpha \right] }\equiv 0$ and $\xi _{%
\left[ \mu \nu |\alpha \right] }\equiv 0$. The generating set of gauge
transformations (\ref{tv7a})--(\ref{tv7b}) is off-shell, second-stage
reducible, the accompanying gauge algebra being obviously Abelian. More
precisely, the gauge transformations (\ref{tv7a}) are off-shell second-stage
reducible because: 1. if in (\ref{tv7a})--(\ref{tv7b}) we make the
transformations
\begin{eqnarray}
\epsilon _{\mu \nu \alpha } &\rightarrow &\epsilon _{\mu \nu \alpha
}^{\left( \omega \right) }=-\frac{1}{2}\partial _{\left[ \mu \right. }\omega
_{\left. \nu \alpha \right] },  \label{tv12} \\
\chi _{\mu \nu |\alpha } &\rightarrow &\chi _{\mu \nu |\alpha }^{\left(
\omega ,\psi \right) }=2\partial _{\alpha }\omega _{\mu \nu }-\partial _{
\left[ \mu \right. }\omega _{\left. \nu \right] \alpha }+\partial _{\left[
\mu \right. }\psi _{\left. \nu \right] \alpha },  \label{tv12a} \\
\xi _{\mu \nu |\alpha } &\rightarrow &\xi _{\mu \nu |\alpha }^{\left(
\varphi \right) }=2\partial _{\alpha }\varphi _{\mu \nu }-\partial _{\left[
\mu \right. }\varphi _{\left. \nu \right] \alpha },  \label{rt3}
\end{eqnarray}%
with $\omega _{\nu \alpha }$ and $\varphi _{\mu \nu }$ antisymmetric and $%
\psi _{\nu \alpha }$ symmetric (but otherwise arbitrary), then the gauge
variations of both tensor fields identically vanish $\delta _{\epsilon
^{\left( \omega \right) },\chi ^{\left( \omega ,\psi \right) }}t_{\lambda
\mu \nu |\alpha }\equiv 0$, $\delta _{\xi _{\mu \nu |\alpha }^{\left(
\varphi \right) }}r_{\mu \nu |\alpha \beta }\equiv 0$; 2. there is no
non-vanishing local transformation of $\varphi _{\mu \nu }$ that
annihilates $\xi _{\mu \nu |\alpha }^{\left( \varphi \right) }$ of the form (%
\ref{rt3}), and hence no further local reducibility identity related with (%
\ref{tv7b}); 3. if in (\ref{tv12})--(\ref{tv12a}) we perform the changes
\begin{equation}
\omega _{\nu \alpha } \rightarrow \omega _{\nu \alpha }^{\left( \theta
\right) }=\partial _{\left[ \nu \right. }\theta _{\left. \alpha \right] },
\qquad
\psi _{\nu \alpha } \rightarrow \psi _{\nu \alpha }^{\left( \theta \right)
}=-3\partial _{\left( \nu \right. }\theta _{\left. \alpha \right) },
\label{tv15a}
\end{equation}%
with $\theta _{\alpha }$ an arbitrary vector field, where $\left( \nu \alpha
\cdots \right) $ signifies symmetrization with respect to the indices
between parentheses without normalization factors, then the transformed
gauge parameters (\ref{tv12})--(\ref{tv12a}) identically vanish $\epsilon
_{\mu \nu \alpha }^{\left( \omega ^{\left( \theta \right) }\right) }\equiv 0$%
, $\chi _{\mu \nu |\alpha }^{\left( \omega ^{\left( \theta \right) },\psi
^{\left( \theta \right) }\right) }\equiv 0$; 4. there is no non-vanishing
local transformation of $\theta _{\nu }$ that simultaneously annihilates $%
\omega _{\nu \alpha }^{\left( \theta \right) }$ and $\psi _{\nu \alpha
}^{\left( \theta \right) }$ of the form (\ref{tv15a}), and
hence no further local reducibility identity related to (\ref{tv7a}).

The construction of the antifield-BRST symmetry for this free theory debuts
with the identification of the algebra on which the BRST differential $s$
acts. The generators of the BRST algebra are of two kinds: fields/ghosts and
antifields. The ghost spectrum for the model under study comprises the
fermionic ghosts $\left\{ \eta _{\lambda \mu \nu },\mathcal{G}_{\mu \nu
|\alpha },\mathcal{C}_{\mu \nu |\alpha }\right\} $ associated with the gauge
parameters $\left\{ \epsilon _{\lambda \mu \nu },\chi _{\mu \nu |\alpha
},\xi _{\mu \nu |\alpha }\right\} $ from (\ref{tv7a})--(\ref{tv7b}), the
bosonic ghosts for ghosts $\left\{ C_{\mu \nu },G_{\nu \alpha },\mathcal{C}%
_{\mu \nu }\right\} $ due to the first-stage reducibility parameters $%
\left\{ \omega _{\mu \nu },\psi _{\nu \alpha },\varphi _{\mu \nu }\right\} $
in (\ref{tv12})--(\ref{rt3}), and also the fermionic ghost for ghost for
ghost $C_{\nu }$ corresponding to the second-stage reducibility parameter $%
\theta _{\nu }$ in (\ref{tv15a}). In order to make compatible
the behavior of the gauge and reducibility parameters with that of the
accompanying ghosts, we ask that $\eta _{\lambda \mu \nu }$, $C_{\mu \nu }$,
and $\mathcal{C}_{\mu \nu }$\ are completely antisymmetric, $\mathcal{G}%
_{\mu \nu |\alpha }$ and $\mathcal{C}_{\mu \nu |\alpha }$ obey the analogue
of the properties fulfilled by the gauge parameters $\chi _{\mu \nu |\alpha
} $ and $\xi _{\mu \nu |\alpha }$, while $G_{\nu \alpha }$ is symmetric. The
antifield spectrum is organized into the antifields $\left\{ t^{\ast \lambda
\mu \nu |\alpha },r^{\ast \mu \nu |\alpha \beta }\right\} $ of the original
tensor fields, together with those of the ghosts, $\left\{ \eta ^{\ast
\lambda \mu \nu },\mathcal{G}^{\ast \mu \nu |\alpha },\mathcal{C}^{\ast \mu
\nu |\alpha }\right\} $, $\left\{ C^{\ast \mu \nu },G^{\ast \nu \alpha },%
\mathcal{C}^{\ast \mu \nu }\right\} $ and respectively $C^{\ast \nu }$, of
statistics opposite to that of the associated field/ghost. It is understood
that $t^{\ast \lambda \mu \nu |\alpha }$ and $r^{\ast \mu \nu |\alpha \beta
} $ exhibit the same mixed-symmetry properties like $t_{\lambda \mu \nu
|\alpha }$ and $r_{\mu \nu |\alpha \beta }$\ and similarly with respect to $%
\eta ^{\ast \lambda \mu \nu }$, $\mathcal{G}^{\ast \mu \nu |\alpha }$, $%
\mathcal{C}^{\ast \mu \nu |\alpha }$, $G^{\ast \nu \alpha }$, $C^{\ast \mu
\nu }$, and $\mathcal{C}^{\ast \nu \alpha }$. For subsequent purposes, we
denote the trace of $t^{\ast \lambda \mu \nu |\alpha }$ by $t^{\ast \lambda
\mu }$, being understood that it is antisymmetric. Similarly, the notation $%
r^{\ast \nu \beta }$ signifies the trace of $r^{\ast \mu \nu |\alpha \beta }$%
, $r^{\ast \nu \beta }=\sigma _{\mu \alpha }r^{\ast \mu \nu |\alpha \beta }$%
, which is symmetric, $r^{\ast \nu \beta }=r^{\ast \beta \nu }$, while $%
r^{\ast }$ denotes its double trace, $r^{\ast }=\sigma ^{\ast \nu \beta
}r_{\nu \beta }\equiv r_{\mu \nu }^{\ast \quad |\mu \nu }$,\ which is a
scalar.

Since both the gauge generators and reducibility functions for this model
are field-independent, it follows that the BRST differential $s$ simply
reduces to
\begin{equation}
s=\delta +\gamma ,  \label{tv39}
\end{equation}%
where $\delta $ represents the Koszul--Tate differential, graded by the
antighost number $\mathrm{agh}$ ($\mathrm{agh}\left( \delta \right) =-1$),
and $\gamma $ stands for the exterior derivative along the gauge orbits,
whose degree is named pure ghost number $\mathrm{pgh}$ ($\mathrm{pgh}\left(
\gamma \right) =1$). These two degrees do not interfere ($\mathrm{agh}\left(
\gamma \right) =0$, $\mathrm{pgh}\left( \delta \right) =0$). The overall
degree that grades the BRST complex is known as the ghost number ($\mathrm{gh%
}$) and is defined like the difference between the pure ghost number and the
antighost number, such that $\mathrm{gh}\left( s\right) =\mathrm{gh}\left(
\delta \right) =\mathrm{gh}\left( \gamma \right) =1$. According to the
standard rules of the BRST method, the corresponding degrees of the
generators from the BRST complex are valued like
\begin{gather}
\mathrm{pgh}\left( t_{\lambda \mu \nu |\alpha }\right) =0=\mathrm{pgh}\left(
r_{\mu \nu |\alpha \beta }\right) ,  \label{rt4} \\
\mathrm{pgh}\left( \eta _{\lambda \mu \nu }\right) =\mathrm{pgh}\left(
\mathcal{G}_{\mu \nu |\alpha }\right) =\mathrm{pgh}\left( \mathcal{C}_{\mu
\nu |\alpha }\right) =1,  \label{rt5} \\
\mathrm{pgh}\left( C_{\mu \nu }\right) =\mathrm{pgh}\left( \mathcal{C}_{\mu
\nu }\right) =\mathrm{pgh}\left( G_{\nu \alpha }\right) =2,\qquad \mathrm{pgh}%
\left( C_{\nu }\right) =3,  \label{rt6} \\
\mathrm{pgh}\left( t^{\ast \lambda \mu \nu |\alpha }\right) =0=\mathrm{pgh}%
\left( r^{\ast \mu \nu |\alpha \beta }\right) ,  \label{rt7} \\
\mathrm{pgh}\left( \eta ^{\ast \lambda \mu \nu }\right) =\mathrm{pgh}\left(
\mathcal{G}^{\ast \mu \nu |\alpha }\right) =\mathrm{pgh}\left( \mathcal{C}%
^{\ast \mu \nu |\alpha }\right) =0,  \label{rt8} \\
\mathrm{pgh}\left( C^{\ast \mu \nu }\right) =\mathrm{pgh}\left( \mathcal{C}%
^{\ast \mu \nu }\right) =\mathrm{pgh}\left( G^{\ast \nu \alpha }\right) =%
\mathrm{pgh}\left( C^{\ast \nu }\right) =0,  \label{rt9} \\
\mathrm{agh}\left( t_{\lambda \mu \nu |\alpha }\right) =0=\mathrm{agh}\left(
r_{\mu \nu |\alpha \beta }\right) ,  \label{rt10} \\
\mathrm{agh}\left( \eta _{\lambda \mu \nu }\right) =\mathrm{agh}\left(
\mathcal{G}_{\mu \nu |\alpha }\right) =\mathrm{agh}\left( \mathcal{C}_{\mu
\nu |\alpha }\right) =0,  \label{rt11} \\
\mathrm{agh}\left( C_{\mu \nu }\right) =\mathrm{agh}\left( \mathcal{C}_{\mu
\nu }\right) =\mathrm{agh}\left( G_{\nu \alpha }\right) =\mathrm{agh}\left(
C_{\nu }\right) =0,  \label{rt12} \\
\mathrm{agh}\left( t^{\ast \lambda \mu \nu |\alpha }\right) =1=\mathrm{agh}%
\left( r^{\ast \mu \nu |\alpha \beta }\right) ,  \label{rt13} \\
\mathrm{agh}\left( \eta ^{\ast \lambda \mu \nu }\right) =\mathrm{agh}\left(
\mathcal{G}^{\ast \mu \nu |\alpha }\right) =\mathrm{agh}\left( \mathcal{C}%
^{\ast \mu \nu |\alpha }\right) =2,  \label{rt14} \\
\mathrm{agh}\left( C^{\ast \mu \nu }\right) =\mathrm{agh}\left( \mathcal{C}%
^{\ast \mu \nu }\right) =\mathrm{agh}\left( G^{\ast \nu \alpha }\right)
=3,\qquad \mathrm{agh}\left( C^{\ast \nu }\right) =4.  \label{rt15}
\end{gather}%
Actually, (\ref{tv39}) is a decomposition of the BRST differential according
to the antighost number and it shows that $s$ contains only components of
antighost number equal to minus one and zero. The Koszul--Tate differential
is imposed to realize a homological resolution of the algebra of smooth
functions defined on the stationary surface of field equations, while the
exterior longitudinal derivative is related to the gauge symmetries (see
relations (\ref{tv7a})--(\ref{tv7b})) of action (\ref{rt1b}) through its
cohomology at pure ghost number zero computed in the cohomology of $\delta $%
, which is required to be the algebra of physical observables for the free
model under consideration. The actions of $\delta $ and $\gamma $ on the
generators from the BRST complex, which enforce all the above mentioned
properties, are given by
\begin{gather}
\gamma t_{\lambda \mu \nu |\alpha }=3\partial _{\alpha }\eta _{\lambda \mu
\nu }+\partial _{\left[ \lambda \right. }\eta _{\left. \mu \nu \right]
\alpha }+\partial _{\left[ \lambda \right. }\mathcal{G}_{\left. \mu \nu %
\right] |\alpha },  \label{rt16} \\
\gamma r_{\mu \nu |\alpha \beta }=\partial _{\mu }\mathcal{C}_{\alpha \beta
|\nu }-\partial _{\nu }\mathcal{C}_{\alpha \beta |\mu }+\partial _{\alpha }%
\mathcal{C}_{\mu \nu |\beta }-\partial _{\beta }\mathcal{C}_{\mu \nu |\alpha
},  \label{rt17} \\
\gamma \eta _{\lambda \mu \nu }=-\frac{1}{2}\partial _{\left[ \lambda
\right. }C_{\left. \mu \nu \right] },\qquad \gamma \mathcal{C}_{\mu \nu
|\alpha }=2\partial _{\alpha }\mathcal{C}_{\mu \nu }-\partial _{\left[ \mu
\right. }\mathcal{C}_{\left. \nu \right] \alpha },  \label{rt18} \\
\gamma \mathcal{G}_{\mu \nu |\alpha }=2\partial _{\left[ \mu \right.
}C_{\left. \nu \alpha \right] }-3\partial _{\left[ \mu \right. }C_{\left.
\nu \right] \alpha }+\partial _{\left[ \mu \right. }G_{\left. \nu \right]
\alpha },  \label{rt19} \\
\gamma C_{\mu \nu }=\partial _{\left[ \mu \right. }C_{\left. \nu \right]
},\qquad \gamma G_{\nu \alpha }=-3\partial _{\left( \nu \right. }C_{\left.
\alpha \right) },\qquad \gamma \mathcal{C}_{\mu \nu }=\gamma C_{\nu }=0,
\label{rt20} \\
\gamma t^{\ast \lambda \mu \nu |\alpha }=\gamma r^{\ast \mu \nu |\alpha
\beta }=\gamma \eta ^{\ast \lambda \mu \nu }=\gamma \mathcal{G}^{\ast \mu
\nu |\alpha }=\gamma \mathcal{C}^{\ast \mu \nu |\alpha }=0,  \label{rt21} \\
\gamma C^{\ast \mu \nu }=\gamma G^{\ast \nu \alpha }=\gamma \mathcal{C}%
^{\ast \mu \nu }=\gamma C^{\ast \nu }=0,  \label{rt22} \\
\delta t_{\lambda \mu \nu |\alpha }=\delta r_{\mu \nu |\alpha \beta }=\delta
\eta _{\lambda \mu \nu }=\delta \mathcal{G}_{\mu \nu |\alpha }=\delta
\mathcal{C}_{\mu \nu |\alpha }=0,  \label{rt23} \\
\delta C_{\mu \nu }=\delta G_{\nu \alpha }=\delta \mathcal{C}_{\mu \nu
}=\delta C_{\nu }=0,  \label{rt24} \\
\delta t^{\ast \lambda \mu \nu |\alpha }=-\frac{\delta S_{0}\left[
t_{\lambda \mu \nu |\alpha }\right] }{\delta t_{\lambda \mu \nu |\alpha }}%
,\qquad \delta r^{\ast \mu \nu |\alpha \beta }=-\frac{\delta S_{0}\left[
r_{\mu \nu |\alpha \beta }\right] }{\delta r_{\mu \nu |\alpha \beta }},
\label{rt25} \\
\delta \eta ^{\ast \lambda \mu \nu }=-4\partial _{\alpha }t^{\ast \lambda
\mu \nu |\alpha },\qquad \delta \mathcal{G}^{\ast \mu \nu |\alpha }=-\partial
_{\lambda }\left( 3t^{\ast \lambda \mu \nu |\alpha }-t^{\ast \mu \nu \alpha
|\lambda }\right) ,  \label{rt26} \\
\delta \mathcal{C}^{\ast \alpha \beta |\nu }=-4\partial _{\mu }r^{\ast \mu
\nu |\alpha \beta },\qquad \delta C^{\ast \mu \nu }=3\partial _{\lambda
}\left( \mathcal{G}^{\ast \mu \nu |\lambda }-\frac{1}{2}\eta ^{\ast \lambda
\mu \nu }\right) ,  \label{rt27} \\
\delta G^{\ast \nu \alpha }=\partial _{\mu }\mathcal{G}^{\ast \mu \left( \nu
|\alpha \right) },\qquad \delta \mathcal{C}^{\ast \mu \nu }=3\partial
_{\alpha }\mathcal{C}^{\ast \mu \nu |\alpha },  \label{rt28} \\
\delta C^{\ast \nu }=6\partial _{\mu }\left( G^{\ast \mu \nu }-\frac{1}{3}%
C^{\ast \mu \nu }\right) .  \label{rt29}
\end{gather}%
By convention, we take $\delta $ and $\gamma $ to act like right
derivations. We note that the action of the Koszul--Tate differential on the
antifields with the antighost number equal to two and respectively three
gains a simpler expression if we perform the changes of variables
\begin{equation}
\mathcal{G}^{\prime \ast \mu \nu |\alpha }=\mathcal{G}^{\ast \mu \nu |\alpha
}+\frac{1}{4}\eta ^{\ast \mu \nu \alpha },\qquad G^{\prime \ast \nu \alpha
}=G^{\ast \nu \alpha }-\frac{1}{3}C^{\ast \nu \alpha }.  \label{tv58a}
\end{equation}%
The antifield $\mathcal{G}^{\prime \ast \mu \nu |\alpha }$ is still
antisymmetric in its first two indices, but does not fulfill any longer the
identity $\mathcal{G}^{\prime \ast \left[ \mu \nu |\alpha \right] }\equiv 0$%
, and $G^{\prime \ast \nu \alpha }$ has no definite symmetry or antisymmetry
properties. With the help of relations (\ref{rt26})--(\ref{rt29}), we find
that $\delta $ acts on the transformed antifields through the relations
\begin{equation}
\delta \mathcal{G}^{\prime \ast \mu \nu |\alpha } =-3\partial _{\lambda
}t^{\ast \lambda \mu \nu |\alpha },  \qquad
\delta G^{\prime \ast \nu \alpha } =2\partial _{\mu }\mathcal{G}^{\prime
\ast \mu \nu |\alpha },  \qquad
\delta C^{\ast \nu } =6\partial _{\mu }G^{\prime \ast \mu \nu }.
\label{tv58b1}
\end{equation}%
The same observation is valid with respect to $\gamma $ if we make the
changes
\begin{equation}
\mathcal{G}_{\mu \nu |\alpha }^{\prime }=\mathcal{G}_{\mu \nu |\alpha
}+4\eta _{\mu \nu \alpha },\qquad G_{\nu \alpha }^{\prime }=G_{\nu \alpha
}-3C_{\nu \alpha },  \label{tv58ba}
\end{equation}%
in terms of which we can write
\begin{equation}
\gamma t_{\lambda \mu \nu |\alpha } =-\frac{1}{4}\partial _{\left[ \lambda
\right. }\mathcal{G}_{\left. \mu \nu |\alpha \right] }^{\prime }+\partial _{%
\left[ \lambda \right. }\mathcal{G}_{\left. \mu \nu \right] |\alpha
}^{\prime },  \qquad
\gamma \mathcal{G}_{\mu \nu |\alpha }^{\prime } =\partial _{\left[ \mu
\right. }G_{\left. \nu \right] \alpha }^{\prime },  \qquad
\gamma G_{\nu \alpha }^{\prime } =-6\partial _{\nu }C_{\alpha }.
\label{tv58bd3}
\end{equation}%
Again, $\mathcal{G}_{\mu \nu |\alpha }^{\prime }$ is antisymmetric in its
first two indices, but does not satisfy the identity $\mathcal{G}_{\left[
\mu \nu |\alpha \right] }^{\prime }\equiv 0$, while $G_{\nu \alpha }^{\prime
}$ has no definite symmetry or antisymmetry. We have deliberately chosen the
same notations for the transformed variables (\ref{tv58a}) and (\ref{tv58ba}%
) since they actually form pairs that are conjugated in the antibracket.

The Lagrangian BRST differential admits a canonical action in a structure
named antibracket and defined by decreeing the fields/ghosts conjugated with
the corresponding antifields, $s\cdot =\left( \cdot ,S\right) $, where $%
\left( ,\right) $ signifies the antibracket and $S$ denotes the canonical
generator of the BRST symmetry. It is a bosonic functional of ghost number
zero, involving both field/ghost and antifield spectra, that obeys the
master equation
\begin{equation}
\left( S,S\right) =0.  \label{tv59}
\end{equation}%
The master equation is equivalent with the second-order nilpotency of $s$,
where its solution $S$ encodes the entire gauge structure of the associated
theory. Taking into account formulae (\ref{rt16})--(\ref{rt29}) as well as
the standard actions of $\delta $ and $\gamma $ in canonical form, we find
that the complete solution to the master equation for the free model under
study is given by
\begin{eqnarray}
S &=&S_{0}\left[ t_{\lambda \mu \nu |\alpha },r_{\mu \nu |\alpha \beta }%
\right] +\int d^{D}x\left[ t^{\ast \lambda \mu \nu |\alpha }\left( 3\partial
_{\alpha }\eta _{\lambda \mu \nu }+\partial _{\left[ \lambda \right. }\eta
_{\left. \mu \nu \right] \alpha }+\partial _{\left[ \lambda \right. }%
\mathcal{G}_{\left. \mu \nu \right] |\alpha }\right) \right.  \notag \\
&&-\frac{1}{2}\eta ^{\ast \mu \nu \alpha }\partial _{\left[ \mu \right.
}C_{\left. \nu \alpha \right] }+\mathcal{G}^{\ast \mu \nu |\alpha }\left(
2\partial _{\alpha }C_{\mu \nu }-\partial _{\left[ \mu \right. }C_{\left.
\nu \right] \alpha }+\partial _{\left[ \mu \right. }G_{\left. \nu \right]
\alpha }\right)  \notag \\
&&+C^{\ast \mu \nu }\partial _{\left[ \mu \right. }C_{\left. \nu \right]
}-3G^{\ast \nu \alpha }\partial _{\left( \nu \right. }C_{\left. \alpha
\right) }+r^{\ast \mu \nu |\alpha \beta }\left( \partial _{\mu }\mathcal{C}%
_{\alpha \beta |\nu }-\partial _{\nu }\mathcal{C}_{\alpha \beta |\mu }\right.
\notag \\
&&\left. \left. +\partial _{\alpha }\mathcal{C}_{\mu \nu |\beta }-\partial
_{\beta }\mathcal{C}_{\mu \nu |\alpha }\right) +\mathcal{C}^{\ast \mu \nu
|\alpha }\left( 2\partial _{\alpha }\mathcal{C}_{\mu \nu }-\partial _{\left[
\mu \right. }\mathcal{C}_{\left. \nu \right] \alpha }\right) \right] ,
\label{tv60}
\end{eqnarray}%
such that it contains pieces with the antighost number ranging from zero to
three.

\section{Brief review of the deformation procedure}

There are three main types of consistent interactions that can be added to a
given gauge theory: \textit{(i)} the first type deforms only the Lagrangian
action, but not its gauge transformations, \textit{(ii)} the second kind
modifies both the action and its transformations, but not the gauge algebra,
and \textit{(iii)} the third, and certainly most interesting category,
changes everything, namely, the action, its gauge symmetries and the
accompanying algebra.

The reformulation of the problem of consistent deformations of a given
action and of its gauge symmetries in the antifield-BRST setting is based on
the observation that if a deformation of the classical theory can be
consistently constructed, then the solution to the master equation for the
initial theory can be deformed into the solution of the master equation for
the interacting theory
\begin{equation}
\bar{S}=S+gS_{1}+g^{2}S_{2}+O\left( g^{3}\right) ,\qquad \varepsilon \left(
\bar{S}\right) =0,\qquad \mathrm{gh}\left( \bar{S}\right) =0,  \label{tv61}
\end{equation}%
such that
\begin{equation}
\left( \bar{S},\bar{S}\right) =0.  \label{tv62}
\end{equation}%
Here and in the sequel $\varepsilon \left( F\right) $ denotes the Grassmann
parity of $F$. The projection of (\ref{tv61}) on the various powers of the
coupling constant induces the following tower of equations:
\begin{eqnarray}
g^{0} &:&\left( S,S\right) =0,  \label{tv63} \\
g^{1} &:&\left( S_{1},S\right) =0,  \label{tv64} \\
g^{2} &:&\frac{1}{2}\left( S_{1},S_{1}\right) +\left( S_{2},S\right) =0,
\label{tv65} \\
g^{3} &:&\left( S_{1},S_{2}\right) +\left( S_{3},S\right) =0,  \label{tv66}
\\
g^{4} &:&\frac{1}{2}\left( S_{2},S_{2}\right) +\left( S_{1},S_{3}\right)
+\left( S_{4},S\right) =0,  \label{tv66a} \\
&&\vdots  \notag
\end{eqnarray}%
The first equation is satisfied by hypothesis. The second one governs the
first-order deformation of the solution to the master equation, $S_{1}$, and
it expresses the fact that $S_{1}$ is a BRST co-cycle, $sS_{1}=0$, and hence
it exists and is local. The remaining equations are responsible for the
higher-order deformations of the solution to the master equation. No
obstructions arise in finding solutions to them as long as no further
restrictions, such as spatiotemporal locality, are imposed. Obviously, only
non-trivial first-order deformations should be considered, since trivial
ones ($S_{1}=sB$) lead to trivial deformations of the initial theory, and
can be eliminated by convenient redefinitions of the fields. Ignoring the
trivial deformations, it follows that $S_{1}$ is a non-trivial
BRST-observable, $S_{1}\in H^{0}\left( s\right) $ (where $H^{0}\left(
s\right) $ denotes the cohomology space of the BRST differential at ghost
number zero). Once the deformation equations ((\ref{tv64})--(\ref{tv66a}),
etc.) have been solved by means of specific cohomological techniques, from
the consistent non-trivial deformed solution to the master equation one can
extract all the information on the gauge structure of the resulting
interacting theory.

\section{Main result\label{mainres}}

The aim of this paper is to investigate the consistent interactions that can
be added to action (\ref{rt1b}) without modifying either the
field/ghost/antifield spectrum or the number of independent gauge
symmetries. This matter is addressed in the context of the antifield-BRST
deformation procedure described in the above and relies on computing the
solutions to Eqs. (\ref{tv64})--(\ref{tv66a}), etc., from the cohomology of
the BRST differential. For obvious reasons, we consider only analytical,
local, and manifestly covariant deformations and, meanwhile, restrict to
Poincar\'{e}-invariant quantities, i.e. we do not allow explicit dependence
on the spatiotemporal coordinates. The analyticity of deformations refers to
the fact that the deformed solution to the master equation, (\ref{tv61}), is
analytical in the coupling constant $g$ and reduces to the original solution
(\ref{tv60}) in the free limit ($g=0$). Moreover, we ask that the deformed
gauge theory preserves the Cauchy order of the uncoupled model, which
enforces the requirement that the interacting Lagrangian is of maximum order
equal to two in the spatiotemporal derivatives of the fields at each order
in the coupling constant. Here, we present the main result without insisting
on the cohomology tools required by the technique of consistent
deformations. All the cohomological proofs that lead to the main result are
given in the next section.

The fully deformed solution to the master equation (\ref{tv62}) ends at
order two in the coupling constant and is given by%
\begin{eqnarray}
\bar{S} &=&S+g\int d^{6}x\,\left[ \varepsilon _{\mu \nu \alpha \lambda \beta
\gamma }\eta ^{\ast \mu \nu \alpha }\partial ^{\lambda }\mathcal{C}^{\beta
\gamma }-2t^{\ast \lambda \mu \nu |\alpha }\varepsilon _{\lambda \mu \nu
\rho \beta \gamma }\left( \partial ^{\rho }\mathcal{C}_{\ \ \ \ \alpha
}^{\beta \gamma |}\right. \right.  \notag \\
&&\left. \left. -\frac{1}{4}\delta _{\ \alpha }^{\gamma }\partial ^{\left[
\rho \right. }\mathcal{C}_{\ \ \ \ \tau }^{\left. \beta \tau \right]
|}\right) -2t_{\lambda \mu \nu |\rho }\varepsilon ^{\lambda \mu \nu \alpha
\beta \gamma }\left( \partial _{\sigma }\partial _{\alpha }r_{\beta \gamma
|}^{\ \ \ \sigma \rho }-\frac{1}{2}\delta _{\ \gamma }^{\rho }\partial
^{\tau }\partial _{\alpha }r_{\beta \tau }\right) +r\right]  \notag \\
&&-g^{2}\int d^{6}x\left( 5r^{\lambda \rho |\left[ \alpha \beta ,\gamma %
\right] }r_{\lambda \rho |\left[ \alpha \beta ,\gamma \right] }-6r_{\lambda
\rho |}^{\ \ \ \left[ \alpha \beta ,\rho \right] }r_{\ \ \ \ \left[ \alpha
\beta ,\sigma \right] }^{\lambda \sigma |}\right) ,  \label{rt30}
\end{eqnarray}%
where%
\begin{equation}
r^{\lambda \rho |\left[ \alpha \beta ,\gamma \right] }=\partial ^{\gamma
}r^{\lambda \rho |\alpha \beta }+\partial ^{\beta }r^{\lambda \rho |\gamma
\alpha }+\partial ^{\alpha }r^{\lambda \rho |\beta \gamma }.  \label{rt31}
\end{equation}%
We recall that $r$ denotes the double trace of $r_{\mu \nu |\alpha \beta }$.
We observe that this solution `lives' in a six-dimensional spacetime. From (%
\ref{rt30}) we read all the information on the gauge structure of the
coupled theory. The terms of antighost number zero in (\ref{rt30}) provide
the Lagrangian action. They can be organized as%
\begin{eqnarray}
\bar{S}_{0}\left[ t_{\lambda \mu \nu |\alpha },r_{\mu \nu |\alpha \beta }%
\right] &=&S_{0}\left[ t_{\lambda \mu \nu |\alpha },r_{\mu \nu |\alpha \beta
}\right] +g\int d^{6}x\left[ r\right.  \notag \\
&&-2t_{\lambda \mu \nu |\rho }\varepsilon ^{\lambda \mu \nu \alpha \beta
\gamma }\left( \partial _{\sigma }\partial _{\alpha }r_{\beta \gamma |}^{\ \
\ \sigma \rho }-\frac{1}{2}\delta _{\ \gamma }^{\rho }\partial ^{\tau
}\partial _{\alpha }r_{\beta \tau }\right) \,  \notag \\
&&\left. -g\left( 5r^{\lambda \rho |\left[ \alpha \beta ,\gamma \right]
}r_{\lambda \rho |\left[ \alpha \beta ,\gamma \right] }-6r_{\lambda \rho
|}^{\ \ \ \left[ \alpha \beta ,\rho \right] }r_{\ \ \ \ \left[ \alpha \beta
,\sigma \right] }^{\lambda \sigma |}\right) \right] ,  \label{rt32}
\end{eqnarray}%
where $S_{0}\left[ t_{\lambda \mu \nu |\alpha },r_{\mu \nu |\alpha \beta }%
\right] $ is the Lagrangian action appearing in (\ref{rt1b}) in $D=6$. We
observe that action (\ref{rt32}) contains only mixing-component terms of
order one and two in the coupling constant. The piece of antighost number
one appearing in (\ref{rt30}) gives the deformed gauge transformations in
the form%
\begin{eqnarray}
\bar{\delta}_{\epsilon ,\chi ,\xi }t_{\lambda \mu \nu |\alpha } &=&3\partial
_{\alpha }\epsilon _{\lambda \mu \nu }+\partial _{\left[ \lambda \right.
}\epsilon _{\left. \mu \nu \right] \alpha }+\partial _{\left[ \lambda
\right. }\chi _{\left. \mu \nu \right] |\alpha }  \notag \\
&&-2g\varepsilon _{\lambda \mu \nu \rho \beta \gamma }\left( \partial ^{\rho
}\xi _{\ \ \ \ \alpha }^{\beta \gamma |}-\frac{1}{4}\delta _{\ \alpha
}^{\gamma }\partial ^{\left[ \rho \right. }\xi _{\ \ \ \ \tau }^{\left.
\beta \tau \right] |}\right) ,  \label{rt33} \\
\bar{\delta}_{\xi }r_{\mu \nu |\alpha \beta } &=&\partial _{\mu }\xi
_{\alpha \beta |\nu }-\partial _{\nu }\xi _{\alpha \beta |\mu }+\partial
_{\alpha }\xi _{\mu \nu |\beta }-\partial _{\beta }\xi _{\mu \nu |\alpha
}=\delta _{\xi }r_{\mu \nu |\alpha \beta }.  \label{rt34}
\end{eqnarray}%
It is interesting to note that only the gauge transformations of the tensor
field $(3,1)$ are modified during the deformation process. This is enforced
at order one in the coupling constant by a term linear in the first-order
derivatives of the gauge parameters from the $(2,2)$ sector. From the terms
of antighost number equal to two present in (\ref{rt30}) we learn that only
the first-order reducibility functions (see (\ref{tv12})) are modified at
order one in the coupling constant, the others coinciding with the original
ones%
\begin{eqnarray}
\epsilon _{\mu \nu \alpha } &\rightarrow &\bar{\epsilon}_{\mu \nu \alpha
}^{\left( \omega ,\varphi \right) }=-\frac{1}{2}\partial _{\left[ \mu
\right. }\omega _{\left. \nu \alpha \right] }+g\varepsilon _{\mu \nu \alpha
\lambda \beta \gamma }\partial ^{\lambda }\varphi ^{\beta \gamma },
\label{rt35} \\
\chi _{\mu \nu |\alpha } &\rightarrow &\chi _{\mu \nu |\alpha }^{\left(
\omega ,\psi \right) }=2\partial _{\alpha }\omega _{\mu \nu }-\partial _{
\left[ \mu \right. }\omega _{\left. \nu \right] \alpha }+\partial _{\left[
\mu \right. }\psi _{\left. \nu \right] \alpha },  \label{rt36} \\
\xi _{\mu \nu |\alpha } &\rightarrow &\xi _{\mu \nu |\alpha }^{\left(
\varphi \right) }=2\partial _{\alpha }\varphi _{\mu \nu }-\partial _{\left[
\mu \right. }\varphi _{\left. \nu \right] \alpha }.  \label{rt37}
\end{eqnarray}%
Consequently, the first-order reducibility relations for $t_{\lambda \mu \nu
|\alpha }$ become
\begin{equation}
\bar{\delta}_{\bar{\epsilon}^{\left( \omega ,\varphi \right) },\chi ^{\left(
\omega ,\psi \right) },\xi ^{\left( \varphi \right) }}t_{\lambda \mu \nu
|\alpha }\equiv 0,  \label{red1deft}
\end{equation}%
while those for $r_{\mu \nu |\alpha \beta }$ are not changed with respect to
the free theory. Since there are no other terms of antighost number two in (%
\ref{rt30}), it follows that the gauge algebra of the coupled model is
unchanged by the deformation procedure, being the same Abelian one like for
the starting free theory. The structure of pieces with the antighost number
equal to three from (\ref{rt30}) implies that the second-order reducibility
functions (\ref{tv15a}) remain the same, and hence the
second-order reducibility relations are exactly the initial ones. It is easy
to see from (\ref{rt32})--(\ref{rt37}) that if we impose the PT-invariance
at the level of the coupled model, then we obtain no interactions (we must
set $g=0$\ in these formulae).

It is important to stress that the problem of obtaining consistent
interactions strongly depends on the spatiotemporal dimension. For instance,
if one starts with action (\ref{rt1b}) in $D>6$, then one inexorably gets $%
\bar{S}=S+g\int d^{D}x\ r$, so no cross-interaction term can be added to
either the original Lagrangian or its gauge transformations.

\section{Proof of the main result}

In the sequel we prove the main result, stated in the previous section.

\subsection{First-order deformation}

If we make the notation $S_{1}=\int d^{D}x\,a$, with $a$ a local function,
then the local form of Eq. (\ref{tv64}), which we have seen that controls
the first-order deformation of the solution to the master equation, becomes
\begin{equation}
sa=\partial _{\mu }m^{\mu },\qquad \mathrm{gh}\left( a\right) =0,\qquad
\varepsilon \left( a\right) =0,  \label{tv65a}
\end{equation}%
for some local $m^{\mu }$, and it shows that the non-integrated density of
the first-order deformation pertains to the local cohomology of $s$ at ghost
number zero, $a\in H^{0}\left( s|d\right) $, where $d$ denotes the exterior
spatiotemporal differential. In order to analyze the above equation, we
develop $a$ according to the antighost number
\begin{equation}
a=\sum\limits_{k=0}^{I}a_{k},\qquad \mathrm{agh}\left( a_{k}\right) =k,\qquad
\mathrm{gh}\left( a_{k}\right) =0,\qquad \varepsilon \left( a_{k}\right) =0,
\label{tv65b}
\end{equation}%
and assume, without loss of generality, that the above decomposition stops
at some finite value of the antighost number, $I$. By taking into account
the splitting (\ref{tv39}) of the BRST differential, Eq. (\ref{tv65a})
becomes equivalent to a tower of local equations, corresponding to the
different decreasing values of the antighost number
\begin{eqnarray}
\gamma a_{I} &=&\partial _{\mu }\overset{(I)}{m}^{\mu },  \label{tv65c} \\
\delta a_{I}+\gamma a_{I-1} &=&\partial _{\mu }\overset{(I-1)}{m}^{\mu },
\label{tv65d} \\
\delta a_{k}+\gamma a_{k-1} &=&\partial _{\mu }\overset{(k-1)}{m}^{\mu
},\qquad I-1\geq k\geq 1,  \label{tv65e}
\end{eqnarray}%
where $\left( \overset{(k)}{m}^{\mu }\right) _{k=\overline{0,I}}$ are some
local currents with $\mathrm{agh}\left( \overset{(k)}{m}^{\mu }\right) =k$.
It can be proved that we can replace Eq. (\ref{tv65c}) at strictly positive
antighost numbers with
\begin{equation}
\gamma a_{I}=0,\qquad \mathrm{agh}\left( a_{I}\right) =I>0.  \label{tv65f}
\end{equation}%
The proof can be done like in the Appendix A, Corollary 1, from \cite%
{noijhep31}. In conclusion, under the assumption that $I>0$, the
representative of highest antighost number from the non-integrated density
of the first-order deformation can always be taken to be $\gamma $-closed,
such that Eq. (\ref{tv65a}) associated with the local form of the
first-order deformation is completely equivalent to the tower of equations
given by (\ref{tv65d})--(\ref{tv65e}) and (\ref{tv65f}).

Before proceeding to the analysis of the solutions to the first-order
deformation equation, let us briefly comment on the uniqueness and
triviality of such solutions. Due to the second-order nilpotency of $\gamma $
($\gamma ^{2}=0$), the solution to the top equation, (\ref{tv65f}), is
clearly unique up to $\gamma $-exact contributions, $a_{I}\rightarrow
a_{I}+\gamma b_{I}$. Meanwhile, if it turns out that $a_{I}$ reduces to $%
\gamma $-exact terms only, $a_{I}=\gamma b_{I}$, then it can be made to
vanish, $a_{I}=0$. In other words, the non-triviality of the first-order
deformation $a$ is translated at its highest antighost number component into
the requirement that $a_{I}\in H^{I}\left( \gamma \right) $, where $%
H^{I}\left( \gamma \right) $ denotes the cohomology of the exterior
longitudinal derivative $\gamma $ at pure ghost number equal to $I$. At the
same time, the general condition on the non-integrated density of the
first-order deformation to be in a non-trivial cohomological class of $%
H^{0}\left( s|d\right) $ shows on the one hand that the solution to Eq. (\ref%
{tv65a}) is unique up to $s$-exact pieces plus total derivatives and, on the
other hand, that if the general solution to (\ref{tv65a}) is completely
trivial, $a=sb+\partial _{\mu }n^{\mu }$, then it can be made to vanish, $%
a=0 $.

\subsubsection{Basic cohomologies}

In the light of the above discussion, now we pass to the investigation of
the solutions to Eqs. (\ref{tv65f}) and (\ref{tv65d})--(\ref{tv65e}). We
have seen that the solution to Eq. (\ref{tv65f}) belongs to the cohomology
of the exterior longitudinal derivative, such that we need to compute $%
H\left( \gamma \right) $ in order to construct the component of highest
antighost number from the first-order deformation. This matter is solved
with the help of definitions (\ref{rt16})--(\ref{rt22}). In order to
determine the cohomology $H(\gamma )$, we split the differential $\gamma $
into two pieces
\begin{equation}
\gamma =\gamma _{\mathrm{t}}+\gamma _{\mathrm{r}},  \label{3.7}
\end{equation}%
where $\gamma _{\mathrm{t}}$ acts non-trivially only on the fields/ghosts
from the $\left( 3,1\right) $ sector, while $\gamma _{\mathrm{r}}$ does the
same thing, but with respect to the $\left( 2,2\right) $ sector. From the
above splitting it follows that the nilpotency of $\gamma $ is equivalent to
the nilpotency and anticommutation of its components
\begin{equation}
\left( \gamma _{\mathrm{t}}\right) ^{2}=0=\left( \gamma _{\mathrm{r}}\right)
^{2},\qquad \gamma _{\mathrm{t}}\gamma _{\mathrm{r}}+\gamma _{\mathrm{r}%
}\gamma _{\mathrm{t}}=0,  \label{3.8}
\end{equation}%
so by K\"{u}nneth's formula we finally find the isomorphism
\begin{equation}
H(\gamma )=H(\gamma _{\mathrm{t}})\otimes H(\gamma _{\mathrm{r}}).
\label{3.9}
\end{equation}%
Using the results from \cite{noijhep31} and \cite{r22,r22th} on the
cohomology of the exterior longitudinal derivative, we can state that $%
H(\gamma )$ is generated on the one hand by $\Theta ^{\ast \Delta }$, $%
K_{\lambda \mu \nu \xi |\alpha \beta }$ and $F_{\mu \nu \lambda |\alpha
\beta \gamma }$ as well as by their spatiotemporal derivatives and, on the
other hand, by the ghosts $\mathcal{F}_{\lambda \mu \nu \alpha }\equiv
\partial _{\left[ \lambda \right. }\eta _{\left. \mu \nu \alpha \right] }$, $%
C_{\nu }$, $\mathcal{C}_{\mu \nu }$\ and $\partial _{\left[ \mu \right. }%
\mathcal{C}_{\left. \nu \alpha \right] }$, where $\Theta ^{\ast \Delta }$
denote all the antifields (from both sectors) and
\begin{eqnarray}
K_{\lambda \mu \nu \xi |\alpha \beta } &=&\partial _{\alpha }\partial _{
\left[ \lambda \right. }t_{\left. \mu \nu \xi \right] |\beta }-\partial
_{\beta }\partial _{\left[ \lambda \right. }t_{\left. \mu \nu \xi \right]
|\alpha },  \label{curvature1} \\
F_{\mu \nu \lambda |\alpha \beta \gamma } &=&\partial _{\lambda }\partial
_{\gamma }r_{\mu \nu |\alpha \beta }+\partial _{\mu }\partial _{\gamma
}r_{\nu \lambda |\alpha \beta }+\partial _{\nu }\partial _{\gamma
}r_{\lambda \mu |\alpha \beta }  \notag \\
&&+\partial _{\lambda }\partial _{\alpha }r_{\mu \nu |\beta \gamma
}+\partial _{\mu }\partial _{\alpha }r_{\nu \lambda |\beta \gamma }+\partial
_{\nu }\partial _{\alpha }r_{\lambda \mu |\beta \gamma }  \notag \\
&&+\partial _{\lambda }\partial _{\beta }r_{\mu \nu |\gamma \alpha
}+\partial _{\mu }\partial _{\beta }r_{\nu \lambda |\gamma \alpha }+\partial
_{\nu }\partial _{\beta }r_{\lambda \mu |\gamma \alpha }  \label{curvature}
\end{eqnarray}%
represent the curvature tensors of $t_{\lambda \mu \nu |\alpha }$ and $%
r_{\mu \nu |\alpha \beta }$. So, the most general, non-trivial
representative from $H\left( \gamma \right) $ for the overall theory (\ref%
{rt1b}) reads as
\begin{equation}
a_{I}=\alpha _{I}\left( \left[ K_{\lambda \mu \nu \xi |\alpha \beta }\right]
,\left[ F_{\mu \nu \lambda |\alpha \beta \gamma }\right] ,\left[ \Theta
^{\ast \Delta }\right] \right) \omega ^{I}\left( \mathcal{F}_{\lambda \mu
\nu \alpha },\mathcal{C}_{\mu \nu },\partial _{\left[ \mu \right. }\mathcal{C%
}_{\left. \nu \alpha \right] },C_{\nu }\right) ,  \label{r79}
\end{equation}%
where the notation $f([q])$ means that $f$ depends on $q$ and its
derivatives up to a finite order, while $\omega ^{I}$ denotes the elements
of pure ghost number $I$ (and antighost number zero) of a basis in the space
of polynomials in the corresponding ghosts and some of their first-order
derivatives. The objects $\alpha _{I}$ (obviously non-trivial in $%
H^{0}\left( \gamma \right) $) were taken to have a bounded number of
derivatives, and therefore they are polynomials in the antifields $\Theta
^{\ast \Delta }$, in the curvature tensors $K_{\lambda \mu \nu \xi |\alpha
\beta }$ and $F_{\mu \nu \lambda |\alpha \beta \gamma }$, as well as in
their derivatives. Due to their $\gamma $-closedness, they are called
invariant polynomials. At zero antighost number, the invariant polynomials
are polynomials in the curvature tensors $K_{\lambda \mu \nu \xi |\alpha
\beta }$ and $F_{\mu \nu \lambda |\alpha \beta \gamma }$ and in their
derivatives.

Replacing the solution (\ref{r79}) into Eq. (\ref{tv65d}) and taking into
account definitions (\ref{rt16})--(\ref{rt22}), we remark that a necessary
(but not sufficient) condition for the existence of (non-trivial) solutions $%
a_{I-1}$ is that the invariant polynomials $\alpha _{I}$ are (non-trivial)
objects from the local cohomology of the Koszul--Tate differential $H\left(
\delta |d\right) $ at antighost number $I>0$ and pure ghost number equal to
zero\footnote{\label{local}We recall that the local cohomology $H\left(
\delta |d\right) $ is completely trivial at both strictly positive antighost
\textit{and} pure ghost numbers (for instance, see~\cite{gen1}, Theorem 5.4
and~\cite{commun1}).}, i.e.,
\begin{equation}
\delta \alpha _{I}=\partial _{\mu }\overset{(I-1)}{j}^{\mu },\qquad \mathrm{%
agh}\left( \overset{(I-1)}{j}^{\mu }\right) =I-1\geq 0,\qquad \mathrm{pgh}%
\left( \overset{(I-1)}{j}^{\mu }\right) =0.  \label{tv82}
\end{equation}%
The above notation is generic, in the sense that $\alpha _{I}$ and $\overset{%
(I-1)}{j}^{\mu }$ may actually carry supplementary Lorentz indices.
Consequently, we need to investigate some of the main properties of the
local cohomology of the Koszul--Tate differential $H\left( \delta |d\right) $
at pure ghost number zero and strictly positive antighost numbers in order
to fully determine the component $a_{I}$ of highest antighost number from
the first-order deformation. As the free model under study is a linear gauge
theory of Cauchy order equal to four, the general results from~\cite%
{gen1,gen11} (see also~\cite{lingr,gen2,multi}) ensure that $H\left( \delta
|d\right) $ (at pure ghost number zero) is trivial at antighost numbers
strictly greater than its Cauchy order
\begin{equation}
H_{I}\left( \delta |d\right) =0,\qquad I>4.  \label{tv83}
\end{equation}%
Moreover, if the invariant polynomial $\alpha _{I}$, with $\mathrm{agh}%
\left( \alpha _{I}\right) =I\geq 4$, is trivial in $H_{I}\left( \delta
|d\right) $, then it can be taken to be trivial also in $H_{I}^{\mathrm{inv}%
}\left( \delta |d\right) $%
\begin{equation}
\left( \alpha _{I}=\delta b_{I+1}+\partial _{\mu }\overset{(I)}{c}^{\mu
},\qquad \mathrm{agh}\left( \alpha _{I}\right) =I\geq 4\right) \Rightarrow
\alpha _{I}=\delta \beta _{I+1}+\partial _{\mu }\overset{(I)}{\gamma }^{\mu
},  \label{tv83a}
\end{equation}%
with $\beta _{I+1}$ and $\overset{(I)}{\gamma }^{\mu }$ invariant
polynomials. (An element of $H_{I}^{\mathrm{inv}}\left( \delta |d\right) $
is defined via an equation similar to (\ref{tv82}), but with the
corresponding current also an invariant polynomial.) The result (\ref{tv83a}%
) can be proved like in the Appendix B, Theorem 3, from \cite{noijhep31}.
This is important since together with (\ref{tv83}) ensures that the entire
local cohomology of the Koszul--Tate differential in the space of invariant
polynomials (characteristic cohomology) is trivial in antighost number
strictly greater than four
\begin{equation}
H_{I}^{\mathrm{inv}}\left( \delta |d\right) =0,\qquad I>4.  \label{tv83b}
\end{equation}%
Looking at the definitions (\ref{tv58b1}) involving the transformed
antifields (\ref{tv58a}) and taking into account formulae (\ref{rt26})--(\ref%
{rt28}) with respect to the $\left( 2,2\right) $ sector, we can organize the
non-trivial representatives of $H_{I}\left( \delta |d\right) $ (at pure
ghost number equal to zero) and $H_{I}^{\mathrm{inv}}\left( \delta
|d\right) $ with $I\geq 2$ in the following table.
\begin{table}[h]
\caption{Non-trivial representatives spanning $H_{I}\left( \delta |d\right) $ and $H_{I}^{\mathrm{inv}}\left( \delta |d\right) $}
\label{tvabledelta}
\begin{center}
\begin{tabular}{@{}cc@{}}
\hline
agh & $H_{I}\left( \delta |d\right) $, $H_{I}^{\mathrm{inv}}\left( \delta |d\right) $\\
\hline
$I>4$ & none \\
$I=4$ & $C^{\ast \mu }$\\
$I=3$ & $G^{\prime \ast \nu \alpha },\mathcal{C}^{\ast \mu \nu }$\\
$I=2$ & $\mathcal{G}^{\prime \ast \mu \nu |\alpha },\mathcal{C}^{\ast \mu \nu |\alpha }$\\
\hline
\end{tabular}
\end{center}
\end{table}

We remark that there is no non-trivial element
in $\left( H_{I}\left( \delta |d\right) \right) _{I\geq 2}$
or $\left( H_{I}^{\mathrm{inv}}\left( \delta |d\right) \right) _{I\geq 2}$
that effectively involves the curvatures $%
K_{\lambda \mu \nu \xi |\alpha \beta }$ and $F_{\mu \nu \lambda |\alpha
\beta \gamma }$ and/or their derivatives, and the same stands for the
quantities that are more than linear in the antifields and/or depend on
their derivatives. In contrast to the groups $\left( H_{I}\left( \delta
|d\right) \right) _{I\geq 2}$ and $\left( H_{I}^{\mathrm{inv}}\left( \delta
|d\right) \right) _{I\geq 2}$, which are finite-dimensional, the cohomology $%
H_{1}\left( \delta |d\right) $ at pure ghost number zero, that is related to
global symmetries and ordinary conservation laws, is infinite-dimensional
since the theory is free.

The previous results on $H\left( \delta |d\right) $ and $H^{\mathrm{inv}%
}\left( \delta |d\right) $ at strictly positive antighost numbers are
important because they control the obstructions to removing the antifields
from the first-order deformation. Indeed, due to (\ref{tv83b}), it follows
that we can successively eliminate all the pieces with $I>4$ from the
non-integrated density of the first-order deformation by adding only trivial
terms (the proof is similar to that from the Appendix C in \cite{noijhep31}%
), so we can take, without loss of non-trivial objects, the condition $I\leq
4$ in the decomposition (\ref{tv65b}). The last representative is of the
form (\ref{r79}), where the invariant polynomial is necessarily a
non-trivial object from $H_{I}^{\mathrm{inv}}\left( \delta |d\right) $ for $%
I=2,3,4$ and respectively from $H_{1}\left( \delta |d\right) $ for $I=1$.

\subsubsection{Computation of first-order deformations}

Now, we have at hand all the necessary ingredients for computing the general
form of the first-order deformation of the solution to the master equation
as solution to Eq. (\ref{tv65a}). In view of this, we decompose the
first-order deformation like%
\begin{equation}
a=a^{\mathrm{t}}+a^{\mathrm{r}}+a^{\mathrm{t-r}},  \label{tv84}
\end{equation}%
where $a^{\mathrm{t}}$ denotes the part responsible for the
self-interactions of the field $t_{\lambda \mu \nu |\alpha }$, $a^{\mathrm{r}%
}$ is related to the self-interactions of the field $r_{\mu \nu |\alpha
\beta }$, and $a^{\mathrm{t-r}}$ signifies the component that describes only
the cross-couplings between $t_{\lambda \mu \nu |\alpha }$ and $r_{\mu \nu
|\alpha \beta }$. Obviously, Eq. (\ref{tv65a}) becomes equivalent with three
equations, one for each component%
\begin{equation}
sa^{\mathrm{t}}=\partial _{\mu }m_{\mathrm{t}}^{\mu },\qquad sa^{\mathrm{r}%
}=\partial _{\mu }m_{\mathrm{r}}^{\mu },\qquad sa^{\mathrm{t-r}}=\partial
_{\mu }m_{\mathrm{t-r}}^{\mu }.  \label{tv85c}
\end{equation}%
The solutions to the first two equations from (\ref{tv85c}) were
investigated in \cite{noijhep31} and respectively \cite{r22} and read as%
\begin{equation}
a^{\mathrm{t}}=0,\qquad a^{\mathrm{r}}=r.  \label{rtn}
\end{equation}

In order to solve the third equation from (\ref{tv85c}), we decompose $a^{%
\mathrm{t-r}}$ along the antighost number like in (\ref{tv65b}) and stop at $%
I=4$%
\begin{equation}
a^{\mathrm{t-r}}=a_{0}^{\mathrm{t-r}}+a_{1}^{\mathrm{t-r}}+a_{2}^{\mathrm{t-r%
}}+a_{3}^{\mathrm{t-r}}+a_{4}^{\mathrm{t-r}},  \label{tv87}
\end{equation}%
where $a_{4}^{\mathrm{t-r}}$ can be taken as solution to the equation $%
\gamma a_{4}^{\mathrm{t-r}}=0$, and therefore it is of the form (\ref{r79})
for $I=4$, with $\alpha _{4}$ an invariant polynomial from $H_{4}^{\mathrm{%
inv}}\left( \delta |d\right) $. Because $H_{4}^{\mathrm{inv}}\left( \delta
|d\right) $ is spanned by $C^{\ast \mu }$ (see Table \ref{tvabledelta}) and $%
a_{4}^{\mathrm{t-r}}$ must yield cross-couplings between $t_{\lambda \mu \nu
|\alpha }$ and $r_{\mu \nu |\alpha \beta }$ with maximum two spatiotemporal
derivatives, it follows that the eligible basis elements at pure ghost
number equal to four remain%
\begin{equation}
\omega ^{4}\left( \mathcal{F}_{\lambda \mu \nu \alpha },\mathcal{C}_{\mu \nu
},\partial _{\left[ \mu \right. }\mathcal{C}_{\left. \nu \alpha \right]
},C_{\nu }\right) :\mathcal{C}_{\alpha \beta }\mathcal{C}_{\lambda \rho },\
\mathcal{C}_{\alpha \beta }\partial _{\left[ \lambda \right. }\mathcal{C}%
_{\left. \rho \sigma \right] }.  \label{rt38}
\end{equation}%
So, up to trivial, $\gamma $-exact contributions, we have that%
\begin{equation}
a_{4}^{\mathrm{t-r}}=C^{\ast \mu }\left( M_{\mu }^{\alpha \beta \lambda \rho
}\mathcal{C}_{\alpha \beta }\mathcal{C}_{\lambda \rho }+N_{\mu }^{\alpha
\beta \lambda \rho \sigma }\mathcal{C}_{\alpha \beta }\partial _{\left[
\lambda \right. }\mathcal{C}_{\left. \rho \sigma \right] }\right) ,
\label{tv88}
\end{equation}%
where $M_{\mu }^{\alpha \beta \lambda \rho }=-M_{\mu }^{\beta \alpha \lambda
\rho }=-M_{\mu }^{\alpha \beta \rho \lambda }=M_{\mu }^{\lambda \rho \alpha
\beta }$ and $N_{\mu }^{\alpha \beta \lambda \rho \sigma }=N_{\mu }^{[\alpha
\beta ]\lambda \rho \sigma }=N_{\mu }^{\alpha \beta \lbrack \lambda \rho
\sigma ]}$ are some non-derivative, real constants. Replacing $a_{4}^{%
\mathrm{t-r}}$ into an equation similar to (\ref{tv65d}) for $I=4$ and
computing $\delta a_{4}^{\mathrm{t-r}}$, it follows that%
\begin{equation}
\delta a_{4}^{\mathrm{t-r}}=\gamma \lambda _{3}+\partial ^{\mu }\tau _{\mu
}-2G^{\prime \ast \nu \mu }\partial _{\left[ \nu \right. }\mathcal{C}%
_{\left. \alpha \beta \right] }\left( 2M_{\mu }^{\alpha \beta \lambda \rho }%
\mathcal{C}_{\lambda \rho }+N_{\mu }^{\alpha \beta \lambda \rho \sigma
}\partial _{\left[ \lambda \right. }\mathcal{C}_{\left. \rho \sigma \right]
}\right) ,  \label{tv89}
\end{equation}%
where%
\begin{eqnarray}
\lambda _{3} &=&-G^{\prime \ast \nu \mu }\left[ 2\mathcal{C}_{\alpha \beta
|\nu }\left( 2M_{\mu }^{\alpha \beta \lambda \rho }\mathcal{C}_{\lambda \rho
}+N_{\mu }^{\alpha \beta \lambda \rho \sigma }\partial _{\left[ \lambda
\right. }\mathcal{C}_{\left. \rho \sigma \right] }\right) \right.  \notag \\
&&\left. +3\mathcal{C}_{\alpha \beta }N_{\mu }^{\alpha \beta \lambda \rho
\sigma }\partial _{\lbrack \lambda }\mathcal{C}_{\rho \sigma ]|\nu }\right] .
\label{rt39}
\end{eqnarray}%
Thus, $a_{3}^{\mathrm{t-r}}$ exists if and only if the third term in the
right-hand side of (\ref{tv89}) can be written in a $\gamma $-exact modulo $%
d $ form%
\begin{equation}
G^{\prime \ast \nu \mu }\partial _{\left[ \nu \right. }\mathcal{C}_{\left.
\alpha \beta \right] }\left( 2M_{\mu }^{\alpha \beta \lambda \rho }\mathcal{C%
}_{\lambda \rho }+N_{\mu }^{\alpha \beta \lambda \rho \sigma }\partial _{%
\left[ \lambda \right. }\mathcal{C}_{\left. \rho \sigma \right] }\right)
=\gamma u_{3}+\partial ^{\mu }\pi _{\mu }.  \label{rt40}
\end{equation}%
Taking the (left) Euler--Lagrange derivative of the above equation with
respect to $G^{\prime \ast \nu \mu }$ and recalling the anticommutativity of
this operation with $\gamma $, we obtain%
\begin{equation}
\partial _{\left[ \nu \right. }\mathcal{C}_{\left. \alpha \beta \right]
}\left( 2M_{\mu }^{\alpha \beta \lambda \rho }\mathcal{C}_{\lambda \rho
}+N_{\mu }^{\alpha \beta \lambda \rho \sigma }\partial _{\left[ \lambda
\right. }\mathcal{C}_{\left. \rho \sigma \right] }\right) =\gamma \left( -%
\frac{\delta ^{L}u_{3}}{\delta G^{\prime \ast \nu \mu }}\right) .
\label{rt41}
\end{equation}%
The last relation shows that the object%
\begin{equation}
\partial _{\left[ \nu \right. }\mathcal{C}_{\left. \alpha \beta \right]
}\left( 2M_{\mu }^{\alpha \beta \lambda \rho }\mathcal{C}_{\lambda \rho
}+N_{\mu }^{\alpha \beta \lambda \rho \sigma }\partial _{\left[ \lambda
\right. }\mathcal{C}_{\left. \rho \sigma \right] }\right) ,  \label{rt42}
\end{equation}%
which is a non-trivial element of $H^{4}\left( \gamma \right) $ (see formula
(\ref{r79})), must be $\gamma $-exact. This takes place if and only if $%
M_{\mu }^{\alpha \beta \lambda \rho }=0=N_{\mu }^{\alpha \beta \lambda \rho
\sigma }$, which further implies%
\begin{equation}
a_{4}^{\mathrm{t-r}}=0,  \label{rt43}
\end{equation}%
and hence the first-order deformation in the cross-coupling sector cannot
end non-trivially at antighost number $I=4$.

Consequently, we pass to $I=3$, in which case we can write%
\begin{equation}
a^{\mathrm{t-r}}=a_{0}^{\mathrm{t-r}}+a_{1}^{\mathrm{t-r}}+a_{2}^{\mathrm{t-r%
}}+a_{3}^{\mathrm{t-r}}.  \label{tv91}
\end{equation}%
Here, $a_{3}^{\mathrm{t-r}}$ is solution to the equation $\gamma a_{3}^{%
\mathrm{t-r}}=0$, and thus is of the type (\ref{r79}) for $I=3$, with $%
\alpha _{3}$ an invariant polynomial from $H_{3}^{\mathrm{inv}}\left( \delta
|d\right) $. There are three independent candidates that comply with all the
hypotheses (including that on the derivative order of the interacting
Lagrangian)%
\begin{equation}
a_{3}^{\mathrm{t-r}}=G^{\prime \ast \nu \alpha }M_{\nu \alpha }^{\lambda
\rho \sigma \tau \gamma \delta }\mathcal{C}_{\lambda \rho }\mathcal{F}%
_{\sigma \tau \gamma \delta }+\mathcal{C}^{\ast \tau \nu }\left( N_{\tau \nu
}^{\alpha }C_{\alpha }+L_{\tau \nu }^{\lambda \rho \sigma \xi \gamma \delta }%
\mathcal{C}_{\lambda \rho }\mathcal{F}_{\sigma \xi \gamma \delta }\right) ,
\label{rt44}
\end{equation}%
where $M_{\nu \alpha }^{\lambda \rho \sigma \tau \gamma \delta }$, $N_{\tau
\nu }^{\alpha }$ and $L_{\tau \nu }^{\lambda \rho \sigma \xi \gamma \delta }$
are some non-derivative, real constants. By direct computation it follows
that%
\begin{eqnarray}
\delta a_{3}^{\mathrm{t-r}} &=&\gamma \lambda _{2}+\partial ^{\mu }\sigma
_{\mu }  \notag \\
&&+\left( \frac{2}{3}\mathcal{G}^{\prime \ast \mu \nu |\alpha }M_{\nu \alpha
}^{\lambda \rho \sigma \xi \gamma \delta }+\mathcal{C}^{\ast \tau \nu |\mu
}L_{\tau \nu }^{\lambda \rho \sigma \xi \gamma \delta }\right) \partial _{%
\left[ \mu \right. }\mathcal{C}_{\left. \lambda \rho \right] }\mathcal{F}%
_{\sigma \xi \gamma \delta },  \label{rt45}
\end{eqnarray}%
with%
\begin{eqnarray}
\lambda _{2} &=&\left( \frac{2}{3}\mathcal{G}^{\prime \ast \mu \nu |\alpha
}M_{\nu \alpha }^{\lambda \rho \sigma \tau \gamma \delta }+\mathcal{C}^{\ast
\xi \nu |\mu }L_{\xi \nu }^{\lambda \rho \sigma \tau \gamma \delta }\right)
\left( -\mathcal{C}_{\lambda \rho |\mu }\mathcal{F}_{\sigma \tau \gamma
\delta }\right.   \notag \\
&&\left. +\partial _{\left[ \sigma \right. }t_{\left. \tau \gamma \delta %
\right] |\mu }\mathcal{C}_{\lambda \rho }\right) -\frac{1}{2}\mathcal{C}%
^{\ast \tau \nu |\mu }N_{\tau \nu }^{\alpha }G_{\mu \alpha }^{\prime }.
\label{rt48}
\end{eqnarray}%
From (\ref{rt45}) we find that $a_{2}^{\mathrm{t-r}}$ exists if and only if
the third term in the right-hand side of (\ref{rt45}) can be written in a $%
\gamma $-exact modulo $d$ form%
\begin{equation}
\left( \frac{2}{3}\mathcal{G}^{\prime \ast \mu \nu |\alpha }M_{\nu \alpha
}^{\lambda \rho \sigma \tau \gamma \delta }+\mathcal{C}^{\ast \xi \nu |\mu
}L_{\xi \nu }^{\lambda \rho \sigma \tau \gamma \delta }\right) \partial _{%
\left[ \mu \right. }\mathcal{C}_{\left. \lambda \rho \right] }\mathcal{F}%
_{\sigma \tau \gamma \delta }=\gamma u_{2}+\partial ^{\mu }l_{\mu }.
\label{rt49}
\end{equation}%
Taking successively the Euler--Lagrange derivatives of (\ref{rt49}) with
respect to the bosonic antifields $\mathcal{G}^{\prime \ast \mu \nu |\alpha }
$ and $\mathcal{C}^{\ast \xi \nu |\mu }$ and taking into account the
commutativity of this operations with $\gamma $, we find%
\begin{equation}
\frac{2}{3}M_{\nu \alpha }^{\lambda \rho \sigma \tau \gamma \delta }\partial
_{\left[ \mu \right. }\mathcal{C}_{\left. \lambda \rho \right] }\mathcal{F}%
_{\sigma \tau \gamma \delta } =\gamma \left( \frac{\delta ^{L}u_{2}}{%
\delta \mathcal{G}^{\prime \ast \mu \nu |\alpha }}\right) ,  \qquad
L_{\xi \nu }^{\lambda \rho \sigma \tau \gamma \delta }\partial _{\left[ \mu
\right. }\mathcal{C}_{\left. \lambda \rho \right] }\mathcal{F}_{\sigma \tau
\gamma \delta } =\gamma \left( \frac{\delta ^{L}u_{2}}{\delta \mathcal{C}%
^{\ast \xi \nu |\mu }}\right) .  \label{rt51}
\end{equation}%
The previous equations indicate that the objects%
\begin{equation}
\frac{2}{3}M_{\nu \alpha }^{\lambda \rho \sigma \tau \gamma \delta }\partial
_{\left[ \mu \right. }\mathcal{C}_{\left. \lambda \rho \right] }\mathcal{F}%
_{\sigma \tau \gamma \delta },\qquad L_{\xi \nu }^{\lambda \rho \sigma \tau
\gamma \delta }\partial _{\left[ \mu \right. }\mathcal{C}_{\left. \lambda
\rho \right] }\mathcal{F}_{\sigma \tau \gamma \delta },  \label{rt52}
\end{equation}%
which are non-trivial elements of $H^{3}\left( \gamma \right) $ (see
relation (\ref{r79})), must be $\gamma $-exact. This takes place if and only
if $M_{\nu \alpha }^{\lambda \rho \sigma \tau \gamma \delta }=0=L_{\xi \nu
}^{\lambda \rho \sigma \tau \gamma \delta }$, which further leads to%
\begin{equation}
a_{3}^{\mathrm{t-r}}=\mathcal{C}^{\ast \tau \nu }N_{\tau \nu }^{\alpha
}C_{\alpha }.  \label{rt53}
\end{equation}%
Then, Eq. (\ref{rt45}) takes the form $\delta a_{3}^{\mathrm{t-r}}=\gamma
\left( -\frac{1}{2}\mathcal{C}^{\ast \tau \nu |\mu }N_{\tau \nu }^{\alpha
}G_{\mu \alpha }^{\prime }\right) +\partial ^{\mu }\sigma _{\mu }$, and
hence
\begin{equation}
a_{2}^{\mathrm{t-r}}=\frac{1}{2}\mathcal{C}^{\ast \tau \nu |\mu }N_{\tau \nu
}^{\alpha }G_{\mu \alpha }^{\prime }+\bar{a}_{2}^{\mathrm{t-r}}.
\label{rt54}
\end{equation}%
In the above $\bar{a}_{2}^{\mathrm{t-r}}$ stands for the general solution to
the homogeneous equation $\gamma \bar{a}_{2}^{\mathrm{t-r}}=0$. It is easy
to see that the only covariant choice of the non-derivative, real constants $%
N_{\tau \nu }^{\alpha }$ is given by $N_{\tau \nu }^{\alpha }=c\varepsilon
_{\quad \tau \nu }^{\alpha }=c\sigma ^{\alpha \beta }\varepsilon _{\beta
\tau \nu }$, where $\varepsilon _{\beta \tau \nu }$ represents the
Levi--Civita symbol in $D=3$ and $c$ is a real constant. Because we work in $%
D\geq 5$, it follows that $c=0$, so $N_{\tau \nu }^{\alpha }=0$. Inserting
this result in formulae (\ref{rt53})--(\ref{rt54}), we then get%
\begin{gather}
a_{3}^{\mathrm{t-r}}=0,\qquad a_{2}^{\mathrm{t-r}}=\bar{a}_{2}^{\mathrm{t-r}},
\label{rt55} \\
\gamma \bar{a}_{2}^{\mathrm{t-r}}=0=\gamma a_{2}^{\mathrm{t-r}}.
\label{rt56}
\end{gather}%
In consequence, the first-order deformation cannot end non-trivially
at antighost number three either.

As a consequence, we can write%
\begin{equation}
a^{\mathrm{t-r}}=a_{0}^{\mathrm{t-r}}+a_{1}^{\mathrm{t-r}}+a_{2}^{\mathrm{t-r%
}},  \label{rt57}
\end{equation}%
with $a_{2}^{\mathrm{t-r}}$ the general solution to the homogeneous equation
$\gamma a_{2}^{\mathrm{t-r}}=0$, and thus of the type (\ref{r79}) for $I=2$,
with $\alpha _{2}$ an invariant polynomial from $H_{2}^{\mathrm{inv}}\left(
\delta |d\right) $. There appear two distinct solutions that fulfill all the
working hypotheses, namely%
\begin{equation}
a_{2}^{\mathrm{t-r}}=\mathcal{G}^{\prime \ast \mu \nu |\alpha }\left( P_{\mu
\nu \alpha }^{\lambda \rho }\mathcal{C}_{\lambda \rho }+Q_{\mu \nu \alpha
}^{\lambda \rho \sigma }\partial _{\left[ \lambda \right. }\mathcal{C}%
_{\left. \rho \sigma \right] }\right) ,  \label{rt58}
\end{equation}%
where $P_{\mu \nu \alpha }^{\lambda \rho }$ and $Q_{\mu \nu \alpha
}^{\lambda \rho \sigma }$ are some non-derivative, real constants, with the
properties $P_{\mu \nu \alpha }^{\lambda \rho }=-P_{\mu \nu \alpha }^{\rho
\lambda }$ and $Q_{\mu \nu \alpha }^{\lambda \rho \sigma }=Q_{\mu \nu \alpha
}^{[\lambda \rho \sigma ]}$. Acting with $\delta $ on (\ref{rt58}), we infer%
\begin{equation}
\delta a_{2}^{\mathrm{t-r}}=\gamma \lambda _{1}+\partial ^{\mu }k_{\mu
}+t^{\ast \tau \mu \nu |\alpha }P_{\mu \nu \alpha }^{\lambda \rho }\partial
_{\left[ \tau \right. }\mathcal{C}_{\left. \lambda \rho \right] },
\label{rt59}
\end{equation}%
where%
\begin{equation}
\lambda _{1}=t^{\ast \tau \mu \nu |\alpha }P_{\mu \nu \alpha }^{\lambda \rho
}\mathcal{C}_{\lambda \rho |\tau }+\frac{3}{2}t^{\ast \tau \mu \nu |\alpha
}Q_{\mu \nu \alpha }^{\lambda \rho \sigma }\partial _{\lbrack \lambda }%
\mathcal{C}_{\rho \sigma ]|\tau }.  \label{rt60}
\end{equation}%
From (\ref{rt59}) we find that $a_{1}^{\mathrm{t-r}}$ exists if and only if
the third term in the right-hand side of (\ref{rt59}) can be written in a $%
\gamma $-exact modulo $d$ form%
\begin{equation}
t^{\ast \tau \mu \nu |\alpha }P_{\mu \nu \alpha }^{\lambda \rho }\partial _{
\left[ \tau \right. }\mathcal{C}_{\left. \lambda \rho \right] }=\gamma
u_{1}+\partial ^{\mu }q_{\mu }.  \label{rt61}
\end{equation}%
Taking the (left) Euler--Lagrange derivative of the above equation with
respect to $t^{\ast \tau \mu \nu |\alpha }$ and recalling the
anticommutativity of this operation with $\gamma $, we deduce%
\begin{equation}
P_{\mu \nu \alpha }^{\lambda \rho }\partial _{\left[ \tau \right. }\mathcal{C%
}_{\left. \lambda \rho \right] }=\gamma \left( -\frac{\delta ^{L}u_{1}}{%
\delta t^{\ast \tau \mu \nu |\alpha }}\right) .  \label{rt62}
\end{equation}%
The previous equation reduces to the requirement that the object%
\begin{equation}
P_{\mu \nu \alpha }^{\lambda \rho }\partial _{\left[ \tau \right. }\mathcal{C%
}_{\left. \lambda \rho \right] },  \label{rt63}
\end{equation}%
which is a non-trivial element of $H^{2}\left( \gamma \right) $ (see
relation (\ref{r79})), must be $\gamma $-exact. This holds if and only if $%
P_{\mu \nu \alpha }^{\lambda \rho }=0$. The last result replaced in formulae
(\ref{rt58})--(\ref{rt60}) yields%
\begin{eqnarray}
a_{2}^{\mathrm{t-r}} &=&\mathcal{G}^{\prime \ast \mu \nu |\alpha }Q_{\mu \nu
\alpha }^{\lambda \rho \sigma }\partial _{\left[ \lambda \right. }\mathcal{C}%
_{\left. \rho \sigma \right] },  \label{rt64} \\
\delta a_{2}^{\mathrm{t-r}} &=&\gamma \left( \frac{3}{2}t^{\ast \tau \mu \nu
|\alpha }Q_{\mu \nu \alpha }^{\lambda \rho \sigma }\partial _{\lbrack
\lambda }\mathcal{C}_{\rho \sigma ]|\tau }\right) +\partial ^{\mu }k_{\mu }.
\label{rt65}
\end{eqnarray}%
Next, Eq. (\ref{rt65}) produces in a simple manner the corresponding $a_{1}^{%
\mathrm{t-r}}$
\begin{equation}
a_{1}^{\mathrm{t-r}}=-\frac{3}{2}t^{\ast \tau \mu \nu |\alpha }Q_{\mu \nu
\alpha }^{\lambda \rho \sigma }\partial _{\lbrack \lambda }\mathcal{C}_{\rho
\sigma ]|\tau }+\bar{a}_{1}^{\mathrm{t-r}},  \label{rt66}
\end{equation}%
where $\bar{a}_{1}^{\mathrm{t-r}}$ means the general solution to the
homogeneous equation $\gamma \bar{a}_{1}^{\mathrm{t-r}}=0$. Recalling the
working hypotheses, we conclude that%
\begin{equation}
\bar{a}_{1}^{\mathrm{t-r}}=r^{\ast \mu \nu |\alpha \beta }Z_{\mu \nu \alpha
\beta }^{\sigma \tau \gamma \delta }\mathcal{F}_{\sigma \tau \gamma \delta },
\label{rt67}
\end{equation}%
where $Z_{\mu \nu \alpha \beta }^{\sigma \tau \gamma \delta }$ denote some
real, non-derivative constants, which are completely antisymmetric with
respect to the indices $\left\{ \sigma ,\tau ,\gamma ,\delta \right\} $. Due
to the mixed symmetry properties of $t^{\ast \mu \nu \alpha |\tau }$ and $%
r^{\ast \mu \nu |\alpha \beta }$, the only covariant choice of $Q_{\mu \nu
\alpha }^{\lambda \rho \sigma }$ and $Z_{\mu \nu \alpha \beta }^{\sigma \tau
\gamma \delta }$ in $D\geq 5$ that does not end up with trivial solutions
reads as%
\begin{equation}
Q_{\mu \nu \alpha }^{\lambda \rho \sigma }=\frac{4}{3}\varepsilon _{\mu \nu
\alpha }^{\quad \ \ \lambda \rho \sigma }=\frac{4}{3}\sigma ^{\lambda
\lambda ^{\prime }}\sigma ^{\rho \rho ^{\prime }}\sigma ^{\sigma \sigma
^{\prime }}\varepsilon _{\mu \nu \alpha \lambda ^{\prime }\rho ^{\prime
}\sigma ^{\prime }},\qquad Z_{\mu \nu \alpha \beta }^{\sigma \tau \gamma
\delta }=0,  \label{rt68}
\end{equation}%
with $\varepsilon _{\mu \nu \alpha \lambda ^{\prime }\rho ^{\prime }\sigma
^{\prime }}$ the six-dimensional Levi--Civita symbol. Inserting (\ref{rt68})
into (\ref{rt64}) and (\ref{rt66})--(\ref{rt67}) and recalling
transformations (\ref{tv58a}), we finally obtain
\begin{eqnarray}
a_{2}^{\mathrm{t-r}} &=&\varepsilon ^{\lambda \mu \nu \alpha \beta \gamma
}\eta _{\lambda \mu \nu }^{\ast }\partial _{\alpha }\mathcal{C}_{\beta
\gamma },  \label{rt69} \\
a_{1}^{\mathrm{t-r}} &=&-2\varepsilon _{\lambda \mu \nu \rho \beta \gamma
}t^{\ast \lambda \mu \nu |\alpha }\left( \partial ^{\rho }\mathcal{C}_{\ \ \
\ \alpha }^{\beta \gamma |}-\frac{1}{4}\delta _{\ \alpha }^{\gamma }\partial
^{\left[ \rho \right. }\mathcal{C}_{\ \ \ \ \tau }^{\left. \beta \tau \right]
|}\right) ,\qquad \bar{a}_{1}^{\mathrm{t-r}}=0.  \label{rt70}
\end{eqnarray}%
The second term from the right-hand side of $a_{1}^{\mathrm{t-r}}$ is
vanishing. Nevertheless, it has been introduced in order to restore the
mixed symmetry $\left( 3,1\right) $ of the quantity $\delta a_{1}^{\mathrm{%
t-r}}/\delta t^{\ast \lambda \mu \nu |\alpha }$. By means of (\ref{rt70}),
we infer%
\begin{equation}
\delta a_{1}^{\mathrm{t-r}}=\gamma \left[ 2\varepsilon ^{\lambda \mu \nu
\alpha \beta \gamma }t_{\lambda \mu \nu |\rho }\left( \partial _{\sigma
}\partial _{\alpha }r_{\beta \gamma |}^{\ \ \ \sigma \rho }-\frac{1}{2}%
\delta _{\ \gamma }^{\rho }\partial ^{\tau }\partial _{\alpha }r_{\beta \tau
}\right) \right] +\partial ^{\mu }p_{\mu },  \label{rt72}
\end{equation}%
which then yields%
\begin{equation}
a_{0}^{\mathrm{t-r}}=-2\varepsilon ^{\lambda \mu \nu \alpha \beta \gamma
}t_{\lambda \mu \nu |\rho }\left( \partial _{\sigma }\partial _{\alpha
}r_{\beta \gamma |}^{\ \ \ \sigma \rho }-\frac{1}{2}\delta _{\ \gamma
}^{\rho }\partial ^{\tau }\partial _{\alpha }r_{\beta \tau }\right) +\bar{a}%
_{0}^{\mathrm{t-r}},  \label{rt73}
\end{equation}%
where $\bar{a}_{0}^{\mathrm{t-r}}$ is the general solution to the
`homogeneous' equation
\begin{equation}
\gamma \bar{a}_{0}^{\mathrm{t-r}}=\partial _{\mu }m^{\mu }.  \label{rt71}
\end{equation}

Next, we investigate the solutions to (\ref{rt71}). There are two main types
of solutions to this equation. The first type, to be denoted by $\bar{a}%
_{0}^{\prime \mathrm{t-r}}$, corresponds to $m^{\mu }=0$ and is given by
gauge-invariant, non-integrated densities constructed out of the original
fields and their spatiotemporal derivatives, which, according to (\ref{r79}%
), are of the form $\bar{a}_{0}^{\prime \mathrm{t-r}}=\bar{a}_{0}^{\prime
\mathrm{t-r}}\left( \left[ K_{\lambda \mu \nu \xi |\alpha \beta }\right] ,%
\left[ F_{\mu \nu \lambda |\alpha \beta \gamma }\right] \right) $, up to the
condition that they effectively describe cross-couplings between the two
types of fields and cannot be written in a divergence-like form. Such a
solution implies at least four derivatives of the fields and consequently
must be forbidden by setting $\bar{a}_{0}^{\prime \mathrm{t-r}}=0$.

The second kind of solutions is associated with $m^{\mu }\neq 0$ in (\ref%
{rt71}), being understood that we discard the divergence-like quantities and
maintain the condition on the maximum derivative order of the interacting
Lagrangian being equal to two. In order to solve this equation we start from
the requirement that $\bar{a}_{0}^{\mathrm{t-r}}$ may contain at most two
derivatives, so it can be decomposed like
\begin{equation}
\bar{a}_{0}^{\mathrm{t-r}}=\omega _{0}+\omega _{1}+\omega _{2},
\label{tww60}
\end{equation}%
where $\left( \omega _{i}\right) _{i=\overline{0,2}}$ contains $i$
derivatives. Due to the different number of derivatives in the components $%
\omega _{0}$, $\omega _{1}$, and $\omega _{2}$, Eq. (\ref{tww60}) is
equivalent to three independent equations
\begin{equation}
\gamma \omega _{k}=\partial _{\mu }j_{k}^{\mu },\qquad k=0,1,2.  \label{twwxy}
\end{equation}

Eq. (\ref{twwxy}) for $k=0$ implies the (necessary) conditions
\begin{equation}
\partial _{\lambda }\left( \frac{\partial \omega _{0}}{\partial t_{\lambda
\mu \nu |\alpha }}\right) =0,\qquad \partial _{\alpha }\left( \frac{\partial
\omega _{0}}{\partial t_{\lambda \mu \nu |\alpha }}\right) =0,\qquad \partial
_{\mu }\left( \frac{\partial \omega _{0}}{\partial r_{\mu \nu |\alpha \beta }%
}\right) =0.  \label{tvcond0}
\end{equation}%
The last equation from (\ref{tvcond0}) possesses only the constant solution
\begin{equation}
\frac{\partial \omega _{0}}{\partial r_{\mu \nu |\alpha \beta }}=k\left(
\sigma ^{\mu \alpha }\sigma ^{\nu \beta }-\sigma ^{\mu \beta }\sigma ^{\nu
\alpha }\right) ,  \label{tvsol0}
\end{equation}%
where $k$ is a real constant, so we find that%
\begin{equation}
\omega _{0}=2kr+B\left( t_{\lambda \mu \nu |\alpha }\right) .  \label{rt75}
\end{equation}%
Since $\omega _{0}$ provides no cross-couplings between $t_{\lambda \mu \nu
|\alpha }$ and $r_{\mu \nu |\alpha \beta }$, we can take
\begin{equation}
\omega _{0}=0  \label{omega0}
\end{equation}%
in (\ref{tww60}).

As a digression, we note that the general solution to the equations
\begin{equation}
\partial _{\lambda }\bar{T}^{\lambda \mu \nu |\alpha }=0,\qquad \partial
_{\alpha }\bar{T}^{\lambda \mu \nu |\alpha }=0  \label{tv32c}
\end{equation}%
(with $\bar{T}^{\lambda \mu \nu |\alpha }$ a covariant tensor field with the
mixed symmetry $\left( 3,1\right) $) reads as \cite{noijhep31}
\begin{equation}
\bar{T}^{\lambda \mu \nu |\alpha }=\partial _{\xi }\partial _{\beta }\bar{%
\Phi}^{\lambda \mu \nu \xi |\alpha \beta },  \label{tv32d}
\end{equation}%
where $\bar{\Phi}^{\rho \lambda \mu \nu |\beta \alpha }$ is a tensor with
the mixed symmetry $(4,2)$. A constant solution $C^{\lambda \mu \nu |\alpha
} $ is excluded from covariance arguments due to the mixed symmetry $\left(
3,1\right) $. Along the same line, the general solution to the equations
\begin{equation}
\partial _{\mu }\bar{R}^{\mu \nu |\alpha \beta }=0  \label{r40a}
\end{equation}%
(with $\bar{R}^{\mu \nu |\alpha \beta }$ a covariant tensor field with the
mixed symmetry $\left( 2,2\right) $) is represented by \cite{r22}
\begin{equation}
\bar{R}^{\mu \nu |\alpha \beta }=\partial _{\rho }\partial _{\gamma }\bar{%
\Omega}^{\mu \nu \rho |\alpha \beta \gamma }+k\left( \sigma ^{\mu \alpha
}\sigma ^{\nu \beta }-\sigma ^{\mu \beta }\sigma ^{\nu \alpha }\right) ,
\label{r40b}
\end{equation}%
where $\bar{\Omega}^{\mu \nu \rho |\alpha \beta \gamma }$ is a tensor with
the mixed symmetry $(3,3)$ and $k$ is an arbitrary, real constant. Now, it
is clear why the solution to the last equation from (\ref{tvcond0}) reduces
to (\ref{tvsol0}): $\partial \omega _{0}/\partial r_{\mu \nu |\alpha \beta }$
displays the mixed symmetry $(2,2)$, but is derivative-free by assumption,
so a term similar to the first one from the right-hand side of (\ref{r40b})
is forbidden.

Eq. (\ref{twwxy}) for $k=1$ leads to the requirements
\begin{equation}
\partial _{\lambda }\left( \frac{\delta \omega _{1}}{\delta t_{\lambda \mu
\nu |\alpha }}\right) =0,\qquad \partial _{\alpha }\left( \frac{\delta \omega
_{1}}{\delta t_{\lambda \mu \nu |\alpha }}\right) =0,\qquad \partial _{\mu
}\left( \frac{\delta \omega _{1}}{\delta r_{\mu \nu |\alpha \beta }}\right)
=0,  \label{tvcond1}
\end{equation}%
where $\delta \omega _{1}/\delta t_{\lambda \mu \nu |\alpha }$ and $\delta
\omega _{1}/\delta r_{\mu \nu |\alpha \beta }$ denote the Euler--Lagrange
derivatives of $\omega _{1}$ with respect to the corresponding fields.
Looking at (\ref{tv32d}) and (\ref{r40b}) and recalling that $\omega _{1}$
is by hypothesis of order one in the spatiotemporal derivatives of the
fields, the only solution to equations (\ref{tvcond1}) reduces to%
\begin{equation}
\frac{\delta \omega _{1}}{\delta r_{\mu \nu |\alpha \beta }}=0=\frac{\delta
\omega _{1}}{\delta t_{\lambda \mu \nu |\alpha }}.  \label{tvsol1}
\end{equation}%
This solution forbids the cross-couplings between the two types of fields,
so we can safely take
\begin{equation}
\omega _{1}=0.  \label{omega1}
\end{equation}

Finally, we pass to Eq. (\ref{twwxy}) for $k=2$, which produces the
restrictions
\begin{equation}
\partial _{\lambda }\left( \frac{\delta \omega _{2}}{\delta t_{\lambda \mu
\nu |\alpha }}\right) =0,\qquad \partial _{\alpha }\left( \frac{\delta \omega
_{2}}{\delta t_{\lambda \mu \nu |\alpha }}\right) =0,\qquad \partial _{\mu
}\left( \frac{\delta \omega _{2}}{\delta r_{\mu \nu |\alpha \beta }}\right)
=0,  \label{tvcond2}
\end{equation}%
with the solutions (see formulae (\ref{tv32d}) and (\ref{r40b}))%
\begin{equation}
\frac{\delta \omega _{2}}{\delta t_{\lambda \mu \nu |\alpha }}=\partial
_{\gamma }\partial _{\sigma }W^{\lambda \mu \nu \gamma |\alpha \sigma
},\qquad \frac{\delta \omega _{2}}{\delta r_{\mu \nu |\alpha \beta }}%
=\partial _{\gamma }\partial _{\sigma }U^{\mu \nu \gamma |\alpha \beta
\sigma }.  \label{tvsol2}
\end{equation}%
The tensor $W^{\lambda \mu \nu \gamma |\alpha \sigma }$ has the mixed
symmetry of the curvature tensor $K^{\lambda \mu \nu \gamma |\alpha \sigma }$
and the tensor $U^{\mu \nu \gamma |\alpha \beta \sigma }$ exhibits the mixed
symmetry of the curvature tensor $F^{\mu \nu \gamma |\alpha \beta \sigma }$.
Both of them are derivative-free since $\omega _{2}$ contains precisely two
derivatives of the fields. At this stage it is useful to introduce a
derivation in the algebra of the fields and of their derivatives that counts
the powers of the fields and of their derivatives
\begin{eqnarray}
N &=&\sum\limits_{k\geq 0}\left[ \left( \partial _{\mu _{1}}\cdots \partial
_{\mu _{k}}t_{\lambda \mu \nu |\alpha }\right) \frac{\partial }{\partial
\left( \partial _{\mu _{1}}\cdots \partial _{\mu _{k}}t_{\lambda \mu \nu
|\alpha }\right) }\right.   \notag \\
&&\left. +\left( \partial _{\mu _{1}}\cdots \partial _{\mu _{k}}r_{\mu \nu
|\alpha \beta }\right) \frac{\partial }{\partial \left( \partial _{\mu
_{1}}\cdots \partial _{\mu _{k}}r_{\mu \nu |\alpha \beta }\right) }\right] ,
\label{tww74}
\end{eqnarray}%
so for every non-integrated density $\rho $ we have that
\begin{equation}
N\rho =t_{\lambda \mu \nu |\alpha }\frac{\delta \rho }{\delta t_{\lambda \mu
\nu |\alpha }}+r_{\mu \nu |\alpha \beta }\frac{\delta \rho }{\delta r_{\mu
\nu |\alpha \beta }}+\partial _{\mu }s^{\mu },  \label{tww75}
\end{equation}%
where $\delta \rho /\delta t_{\mu \nu |\alpha \beta }$ and $\delta \rho
/\delta r_{\mu \nu |\alpha \beta }$ denote the variational derivatives of $%
\rho $ with respect to the fields. If $\rho ^{\left( l\right) }$ is a
homogeneous polynomial of order $l>0$ in the fields $\left\{ t_{\lambda \mu
\nu |\alpha },r_{\mu \nu |\alpha \beta }\right\} $ and their derivatives,
then $N\rho ^{\left( l\right) }=l\rho ^{\left( l\right) }$. Using (\ref%
{tvsol2}) and (\ref{tww75}), we find that
\begin{equation}
N\omega _{2}=\frac{1}{8}K_{\lambda \mu \nu \gamma |\alpha \sigma }W^{\lambda
\mu \nu \gamma |\alpha \sigma }+\frac{1}{9}F_{\mu \nu \gamma |\alpha \beta
\sigma }U^{\mu \nu \gamma |\alpha \beta \sigma }+\partial _{\mu }v^{\mu }.
\label{tww76a}
\end{equation}%
We expand $\omega _{2}$ according to the various eigenvalues of $N$ like%
\begin{equation}
\omega _{2}=\sum\limits_{l>0}\omega _{2}^{\left( l\right) },  \label{tww77}
\end{equation}%
where $N\omega _{2}^{\left( l\right) }=l\omega _{2}^{\left( l\right) }$,
such that
\begin{equation}
N\omega _{2}=\sum\limits_{l>0}l\omega _{2}^{\left( l\right) }.  \label{tww78}
\end{equation}%
Comparing (\ref{tww76a}) with (\ref{tww78}), we reach the conclusion that
the decomposition (\ref{tww77}) induces a similar decomposition with respect
to $W^{\lambda \mu \nu \gamma |\alpha \sigma }$ and $U^{\mu \nu \gamma
|\alpha \beta \sigma }$
\begin{equation}
W^{\lambda \mu \nu \gamma |\alpha \sigma }=\sum\limits_{l>0}W_{\left(
l-1\right) }^{\lambda \mu \nu \gamma |\alpha \sigma },\qquad U^{\mu \nu
\gamma |\alpha \beta \sigma }=\sum\limits_{l>0}U_{\left( l-1\right) }^{\mu
\nu \gamma |\alpha \beta \sigma }.  \label{tww79}
\end{equation}%
Substituting (\ref{tww79}) into (\ref{tww76a}) and comparing the resulting
expression with (\ref{tww78}), we obtain that
\begin{equation}
\omega _{2}^{\left( l\right) }=\frac{1}{8l}K_{\lambda \mu \nu \gamma |\alpha
\sigma }W_{\left( l-1\right) }^{\lambda \mu \nu \gamma |\alpha \sigma }+%
\frac{1}{9l}F_{\mu \nu \gamma |\alpha \beta \sigma }U_{\left( l-1\right)
}^{\mu \nu \gamma |\alpha \beta \sigma }+\partial _{\mu }\bar{v}_{(l)}^{\mu
}.  \label{tvprform}
\end{equation}%
Introducing (\ref{tvprform}) in (\ref{tww77}), we arrive at
\begin{equation}
\omega _{2}=K_{\lambda \mu \nu \gamma |\alpha \sigma }\bar{W}^{\lambda \mu
\nu \gamma |\alpha \sigma }+F_{\mu \nu \gamma |\alpha \beta \sigma }\bar{U}%
^{\mu \nu \gamma |\alpha \beta \sigma }+\partial _{\mu }\bar{v}^{\mu },
\label{tww81}
\end{equation}%
where
\begin{equation}
\bar{W}^{\lambda \mu \nu \gamma |\alpha \sigma }=\sum\limits_{l>0}\frac{1}{8l%
}W_{\left( l-1\right) }^{\lambda \mu \nu \gamma |\alpha \sigma },\qquad \bar{U%
}^{\mu \nu \gamma |\alpha \beta \sigma }=\sum\limits_{l>0}\frac{1}{9l}%
U_{\left( l-1\right) }^{\mu \nu \gamma |\alpha \beta \sigma }.  \label{tww82}
\end{equation}%
Applying $\gamma $ on (\ref{tww81}), we infer that a necessary condition for
the existence of solutions to the equation $\gamma \omega _{2}=\partial
_{\mu }j_{2}^{\mu }$ is that the functions $\bar{W}^{\lambda \mu \nu \gamma
|\alpha \sigma }$ and $\bar{U}^{\mu \nu \gamma |\alpha \beta \sigma }$
entering (\ref{tww81}) must satisfy the equations
\begin{eqnarray}
\partial _{\rho }\left( F_{\mu \nu \gamma |\alpha \beta \sigma }\frac{%
\partial \bar{U}^{\mu \nu \gamma |\alpha \beta \sigma }}{\partial r_{\rho
\delta |\xi \chi }}+K_{\lambda \mu \nu \gamma |\alpha \sigma }\frac{\partial
\bar{W}^{\lambda \mu \nu \gamma |\alpha \sigma }}{\partial r_{\rho \delta
|\xi \chi }}\right)  &=&0,  \label{rt80} \\
\partial _{\chi }\left( F_{\mu \nu \gamma |\alpha \beta \sigma }\frac{%
\partial \bar{U}^{\mu \nu \gamma |\alpha \beta \sigma }}{\partial t_{\rho
\delta \xi |\chi }}+K_{\lambda \mu \nu \gamma |\alpha \sigma }\frac{\partial
\bar{W}^{\lambda \mu \nu \gamma |\alpha \sigma }}{\partial t_{\rho \delta
\xi |\chi }}\right)  &=&0,  \label{rt81} \\
\partial _{\rho }\left( F_{\mu \nu \gamma |\alpha \beta \sigma }\frac{%
\partial \bar{U}^{\mu \nu \gamma |\alpha \beta \sigma }}{\partial t_{\rho
\delta \xi |\chi }}+K_{\lambda \mu \nu \gamma |\alpha \sigma }\frac{\partial
\bar{W}^{\lambda \mu \nu \gamma |\alpha \sigma }}{\partial t_{\rho \delta
\xi |\chi }}\right)  &=&0.  \label{rt82}
\end{eqnarray}%
The general solution to Eqs. (\ref{rt80})--(\ref{rt82}) reads as%
\begin{eqnarray}
F_{\mu \nu \gamma |\alpha \beta \sigma }\frac{\partial \bar{U}^{\mu \nu
\gamma |\alpha \beta \sigma }}{\partial r_{\rho \delta |\xi \chi }}%
+K_{\lambda \mu \nu \gamma |\alpha \sigma }\frac{\partial \bar{W}^{\lambda
\mu \nu \gamma |\alpha \sigma }}{\partial r_{\rho \delta |\xi \chi }}
&=&\partial _{\tau }\partial _{\theta }E^{\rho \delta \tau |\xi \chi \theta
},  \label{rt83} \\
F_{\mu \nu \gamma |\alpha \beta \sigma }\frac{\partial \bar{U}^{\mu \nu
\gamma |\alpha \beta \sigma }}{\partial t_{\rho \delta \xi |\chi }}%
+K_{\lambda \mu \nu \gamma |\alpha \sigma }\frac{\partial \bar{W}^{\lambda
\mu \nu \gamma |\alpha \sigma }}{\partial t_{\rho \delta \xi |\chi }}
&=&\partial _{\tau }\partial _{\theta }H^{\rho \delta \xi \tau |\chi \theta
},  \label{rt84}
\end{eqnarray}%
where the functions $E^{\rho \delta \tau |\xi \chi \theta }$ and $H^{\rho
\delta \xi \tau |\chi \theta }$ are derivative-free and exhibit the mixed
symmetries $(3,3)$ and $(4,2)$ respectively. By direct computations we deduce%
\begin{eqnarray}
\partial _{\tau }\partial _{\theta }E^{\rho \delta \tau |\xi \chi \theta }
&=&\frac{\partial ^{2}E^{\rho \delta \tau |\xi \chi \theta }}{\partial
r_{\rho ^{\prime }\delta ^{\prime }|\xi ^{\prime }\chi ^{\prime }}\partial
r_{\rho ^{\prime \prime }\delta ^{\prime \prime }|\xi ^{\prime \prime }\chi
^{\prime \prime }}}\left( \partial _{\theta }r_{\rho ^{\prime }\delta
^{\prime }|\xi ^{\prime }\chi ^{\prime }}\right) \left( \partial _{\tau
}r_{\rho ^{\prime \prime }\delta ^{\prime \prime }|\xi ^{\prime \prime }\chi
^{\prime \prime }}\right)   \notag \\
&&+\frac{\partial ^{2}E^{\rho \delta \tau |\xi \chi \theta }}{\partial
t_{\rho ^{\prime }\delta ^{\prime }\xi ^{\prime }|\chi ^{\prime }}\partial
t_{\rho ^{\prime \prime }\delta ^{\prime \prime }\xi ^{\prime \prime }|\chi
^{\prime \prime }}}\left( \partial _{\theta }t_{\rho ^{\prime }\delta
^{\prime }\xi ^{\prime }|\chi ^{\prime }}\right) \left( \partial _{\tau
}t_{\rho ^{\prime \prime }\delta ^{\prime \prime }\xi ^{\prime \prime }|\chi
^{\prime \prime }}\right)   \notag \\
&&+\frac{\partial ^{2}\left( E^{\rho \delta \tau |\xi \chi \theta }+E^{\rho
\delta \theta |\xi \chi \tau }\right) }{\partial r_{\rho ^{\prime }\delta
^{\prime }|\xi ^{\prime }\chi ^{\prime }}\partial t_{\rho ^{\prime \prime
}\delta ^{\prime \prime }\xi ^{\prime \prime }|\chi ^{\prime \prime }}}%
\left( \partial _{\theta }r_{\rho ^{\prime }\delta ^{\prime }|\xi ^{\prime
}\chi ^{\prime }}\right) \left( \partial _{\tau }t_{\rho ^{\prime \prime
}\delta ^{\prime \prime }\xi ^{\prime \prime }|\chi ^{\prime \prime
}}\right)   \notag \\
&&+\frac{\partial E^{\rho \delta \tau |\xi \chi \theta }}{\partial r_{\rho
^{\prime }\delta ^{\prime }|\xi ^{\prime }\chi ^{\prime }}}\partial _{\tau
}\partial _{\theta }r_{\rho ^{\prime }\delta ^{\prime }|\xi ^{\prime }\chi
^{\prime }}+\frac{\partial E^{\rho \delta \tau |\xi \chi \theta }}{\partial
t_{\rho ^{\prime }\delta ^{\prime }\xi ^{\prime }|\chi ^{\prime }}}\partial
_{\tau }\partial _{\theta }t_{\rho ^{\prime }\delta ^{\prime }\xi ^{\prime
}|\chi ^{\prime }},  \label{rt88}
\end{eqnarray}%
\begin{eqnarray}
\partial _{\tau }\partial _{\theta }H^{\rho \delta \xi \tau |\chi \theta }
&=&\frac{\partial ^{2}H^{\rho \delta \xi \tau |\chi \theta }}{\partial
r_{\rho ^{\prime }\delta ^{\prime }|\xi ^{\prime }\chi ^{\prime }}\partial
r_{\rho ^{\prime \prime }\delta ^{\prime \prime }|\xi ^{\prime \prime }\chi
^{\prime \prime }}}\left( \partial _{\theta }r_{\rho ^{\prime }\delta
^{\prime }|\xi ^{\prime }\chi ^{\prime }}\right) \left( \partial _{\tau
}r_{\rho ^{\prime \prime }\delta ^{\prime \prime }|\xi ^{\prime \prime }\chi
^{\prime \prime }}\right)   \notag \\
&&+\frac{\partial ^{2}H^{\rho \delta \xi \tau |\chi \theta }}{\partial
t_{\rho ^{\prime }\delta ^{\prime }\xi ^{\prime }|\chi ^{\prime }}\partial
t_{\rho ^{\prime \prime }\delta ^{\prime \prime }\xi ^{\prime \prime }|\chi
^{\prime \prime }}}\left( \partial _{\theta }t_{\rho ^{\prime }\delta
^{\prime }\xi ^{\prime }|\chi ^{\prime }}\right) \left( \partial _{\tau
}t_{\rho ^{\prime \prime }\delta ^{\prime \prime }\xi ^{\prime \prime }|\chi
^{\prime \prime }}\right)   \notag \\
&&+\frac{\partial ^{2}\left( H^{\rho \delta \xi \tau |\chi \theta }+H^{\rho
\delta \xi \theta |\chi \tau }\right) }{\partial r_{\rho ^{\prime }\delta
^{\prime }|\xi ^{\prime }\chi ^{\prime }}\partial t_{\rho ^{\prime \prime
}\delta ^{\prime \prime }\xi ^{\prime \prime }|\chi ^{\prime \prime }}}%
\left( \partial _{\theta }r_{\rho ^{\prime }\delta ^{\prime }|\xi ^{\prime
}\chi ^{\prime }}\right) \left( \partial _{\tau }t_{\rho ^{\prime \prime
}\delta ^{\prime \prime }\xi ^{\prime \prime }|\chi ^{\prime \prime
}}\right)   \notag \\
&&+\frac{\partial H^{\rho \delta \xi \tau |\chi \theta }}{\partial r_{\rho
^{\prime }\delta ^{\prime }|\xi ^{\prime }\chi ^{\prime }}}\partial _{\tau
}\partial _{\theta }r_{\rho ^{\prime }\delta ^{\prime }|\xi ^{\prime }\chi
^{\prime }}+\frac{\partial H^{\rho \delta \xi \tau |\chi \theta }}{\partial
t_{\rho ^{\prime }\delta ^{\prime }\xi ^{\prime }|\chi ^{\prime }}}\partial
_{\tau }\partial _{\theta }t_{\rho ^{\prime }\delta ^{\prime }\xi ^{\prime
}|\chi ^{\prime }}.  \label{rt89}
\end{eqnarray}%
Substituting (\ref{rt88})--(\ref{rt89}) in (\ref{rt83})--(\ref{rt84}) and
comparing the left-hand sides with the corresponding right-hand sides of the
resulting relations, we find the necessary equations%
\begin{eqnarray}
\frac{\partial ^{2}E^{\rho \delta \tau |\xi \chi \theta }}{\partial r_{\rho
^{\prime }\delta ^{\prime }|\xi ^{\prime }\chi ^{\prime }}\partial r_{\rho
^{\prime \prime }\delta ^{\prime \prime }|\xi ^{\prime \prime }\chi ^{\prime
\prime }}} &=&0,\qquad \frac{\partial ^{2}E^{\rho \delta \tau |\xi \chi
\theta }}{\partial t_{\rho ^{\prime }\delta ^{\prime }\xi ^{\prime }|\chi
^{\prime }}\partial t_{\rho ^{\prime \prime }\delta ^{\prime \prime }\xi
^{\prime \prime }|\chi ^{\prime \prime }}}=0,  \label{rt90} \\
\frac{\partial ^{2}H^{\rho \delta \xi \tau |\chi \theta }}{\partial r_{\rho
^{\prime }\delta ^{\prime }|\xi ^{\prime }\chi ^{\prime }}\partial r_{\rho
^{\prime \prime }\delta ^{\prime \prime }|\xi ^{\prime \prime }\chi ^{\prime
\prime }}} &=&0,\qquad \frac{\partial ^{2}H^{\rho \delta \xi \tau |\chi
\theta }}{\partial t_{\rho ^{\prime }\delta ^{\prime }\xi ^{\prime }|\chi
^{\prime }}\partial t_{\rho ^{\prime \prime }\delta ^{\prime \prime }\xi
^{\prime \prime }|\chi ^{\prime \prime }}}=0,  \label{rt91} \\
\frac{\partial ^{2}\left( E^{\rho \delta \tau |\xi \chi \theta }+E^{\rho
\delta \theta |\xi \chi \tau }\right) }{\partial r_{\rho ^{\prime }\delta
^{\prime }|\xi ^{\prime }\chi ^{\prime }}\partial t_{\rho ^{\prime \prime
}\delta ^{\prime \prime }\xi ^{\prime \prime }|\chi ^{\prime \prime }}}
&=&0,\qquad \frac{\partial ^{2}\left( H^{\rho \delta \xi \tau |\chi \theta
}+H^{\rho \delta \xi \theta |\chi \tau }\right) }{\partial r_{\rho ^{\prime
}\delta ^{\prime }|\xi ^{\prime }\chi ^{\prime }}\partial t_{\rho ^{\prime
\prime }\delta ^{\prime \prime }\xi ^{\prime \prime }|\chi ^{\prime \prime }}%
}=0.  \label{rt92}
\end{eqnarray}%
The above relations allow us to write%
\begin{equation}
\frac{1}{2}\left( E^{\rho \delta \tau |\xi \chi \theta }+E^{\rho \delta
\theta |\xi \chi \tau }\right) =C^{\rho \delta \tau |\xi \chi \theta ;\rho
^{\prime }\delta ^{\prime }|\xi ^{\prime }\chi ^{\prime }}r_{\rho ^{\prime
}\delta ^{\prime }|\xi ^{\prime }\chi ^{\prime }}+C^{\rho \delta \theta |\xi
\chi \tau ;\rho ^{\prime }\delta ^{\prime }\xi ^{\prime }|\chi ^{\prime
}}t_{\rho ^{\prime }\delta ^{\prime }\xi ^{\prime }|\chi ^{\prime }},
\label{rt93}
\end{equation}%
\begin{equation}
\frac{1}{2}\left( H^{\rho \delta \xi \tau |\chi \theta }+H^{\rho \delta \xi
\theta |\chi \tau }\right) =\hat{C}^{\rho \delta \xi \tau |\chi \theta ;\rho
^{\prime }\delta ^{\prime }|\xi ^{\prime }\chi ^{\prime }}r_{\rho ^{\prime
}\delta ^{\prime }|\xi ^{\prime }\chi ^{\prime }}+\hat{C}^{\rho \delta \xi
\tau |\chi \theta ;\rho ^{\prime }\delta ^{\prime }\xi ^{\prime }|\chi
^{\prime }}t_{\rho ^{\prime }\delta ^{\prime }\xi ^{\prime }|\chi ^{\prime
}},  \label{rt94}
\end{equation}%
where the quantities denoted by $C$ or $\hat{C}$ are some non-derivative,
real tensors, with the expressions%
\begin{eqnarray}
C^{\rho \delta \tau |\xi \chi \theta ;\rho ^{\prime }\delta ^{\prime }|\xi
^{\prime }\chi ^{\prime }} &=&\tilde{C}^{\rho \delta \tau |\xi \chi \theta
;\rho ^{\prime }\delta ^{\prime }|\xi ^{\prime }\chi ^{\prime }}+\tilde{C}%
^{\rho \delta \theta |\xi \chi \tau ;\rho ^{\prime }\delta ^{\prime }|\xi
^{\prime }\chi ^{\prime }},  \label{rt95} \\
C^{\rho \delta \theta |\xi \chi \tau ;\rho ^{\prime }\delta ^{\prime }\xi
^{\prime }|\chi ^{\prime }} &=&\tilde{C}^{\rho \delta \tau |\xi \chi \theta
;\rho ^{\prime }\delta ^{\prime }\xi ^{\prime }|\chi ^{\prime }}+\tilde{C}%
^{\rho \delta \theta |\xi \chi \tau ;\rho ^{\prime }\delta ^{\prime }\xi
^{\prime }|\chi ^{\prime }},  \label{rt96} \\
\hat{C}^{\rho \delta \xi \tau |\chi \theta ;\rho ^{\prime }\delta ^{\prime
}|\xi ^{\prime }\chi ^{\prime }} &=&\bar{C}^{\rho \delta \xi \tau |\chi
\theta ;\rho ^{\prime }\delta ^{\prime }|\xi ^{\prime }\chi ^{\prime }}+\bar{%
C}^{\rho \delta \xi \theta |\chi \tau ;\rho ^{\prime }\delta ^{\prime }|\xi
^{\prime }\chi ^{\prime }},  \label{rt97} \\
\hat{C}^{\rho \delta \xi \tau |\chi \theta ;\rho ^{\prime }\delta ^{\prime
}\xi ^{\prime }|\chi ^{\prime }} &=&\bar{C}^{\rho \delta \xi \tau |\chi
\theta ;\rho ^{\prime }\delta ^{\prime }\xi ^{\prime }|\chi ^{\prime }}+\bar{%
C}^{\rho \delta \xi \theta |\chi \tau ;\rho ^{\prime }\delta ^{\prime }\xi
^{\prime }|\chi ^{\prime }}.  \label{rt98}
\end{eqnarray}%
Wherever two sets of indices are connected by a semicolon, it is understood
that the corresponding tensor possesses independently the mixed symmetries
with respect to the former and respectively the latter set. On the other
hand, it is obvious that%
\begin{eqnarray}
\partial _{\tau }\partial _{\theta }E^{\rho \delta \tau |\xi \chi \theta }
&=&\frac{1}{2}\partial _{\tau }\partial _{\theta }\left( E^{\rho \delta \tau
|\xi \chi \theta }+E^{\rho \delta \theta |\xi \chi \tau }\right) ,
\label{rt99} \\
\partial _{\tau }\partial _{\theta }H^{\rho \delta \xi \tau |\chi \theta }
&=&\frac{1}{2}\partial _{\tau }\partial _{\theta }\left( H^{\rho \delta \xi
\tau |\chi \theta }+H^{\rho \delta \xi \theta |\chi \tau }\right) ,
\label{rt100}
\end{eqnarray}%
so Eqs. (\ref{rt83})--(\ref{rt84}) become%
\begin{gather}
F_{\mu \nu \gamma |\alpha \beta \sigma }\frac{\partial \bar{U}^{\mu \nu
\gamma |\alpha \beta \sigma }}{\partial r_{\rho \delta |\xi \chi }}%
+K_{\lambda \mu \nu \gamma |\alpha \sigma }\frac{\partial \bar{W}^{\lambda
\mu \nu \gamma |\alpha \sigma }}{\partial r_{\rho \delta |\xi \chi }}=
\notag \\
=C^{\rho \delta \tau |\xi \chi \theta ;\rho ^{\prime }\delta ^{\prime }|\xi
^{\prime }\chi ^{\prime }}\partial _{\tau }\partial _{\theta }r_{\rho
^{\prime }\delta ^{\prime }|\xi ^{\prime }\chi ^{\prime }}+C^{\rho \delta
\theta |\xi \chi \tau ;\rho ^{\prime }\delta ^{\prime }\xi ^{\prime }|\chi
^{\prime }}\partial _{\tau }\partial _{\theta }t_{\rho ^{\prime }\delta
^{\prime }\xi ^{\prime }|\chi ^{\prime }},  \label{rt101} \\
F_{\mu \nu \gamma |\alpha \beta \sigma }\frac{\partial \bar{U}^{\mu \nu
\gamma |\alpha \beta \sigma }}{\partial t_{\rho \delta \xi |\chi }}%
+K_{\lambda \mu \nu \gamma |\alpha \sigma }\frac{\partial \bar{W}^{\lambda
\mu \nu \gamma |\alpha \sigma }}{\partial t_{\rho \delta \xi |\chi }}=
\notag \\
=\hat{C}^{\rho \delta \xi \tau |\chi \theta ;\rho ^{\prime }\delta ^{\prime
}|\xi ^{\prime }\chi ^{\prime }}\partial _{\tau }\partial _{\theta }r_{\rho
^{\prime }\delta ^{\prime }|\xi ^{\prime }\chi ^{\prime }}+\hat{C}^{\rho
\delta \xi \tau |\chi \theta ;\rho ^{\prime }\delta ^{\prime }\xi ^{\prime
}|\chi ^{\prime }}\partial _{\tau }\partial _{\theta }t_{\rho ^{\prime
}\delta ^{\prime }\xi ^{\prime }|\chi ^{\prime }}.  \label{rt102}
\end{gather}%
Taking the partial derivatives of Eqs. (\ref{rt101}) and (\ref{rt102}) with
respect to $\partial _{\tau }\partial _{\theta }r_{\rho ^{\prime }\delta
^{\prime }|\xi ^{\prime }\chi ^{\prime }}$ and $\partial _{\tau }\partial
_{\theta }t_{\rho ^{\prime }\delta ^{\prime }\xi ^{\prime }|\chi ^{\prime }}$%
, we infer the relations%
\begin{eqnarray}
\frac{\partial \bar{U}^{\mu \nu \gamma |\alpha \beta \sigma }}{\partial
r_{\rho \delta |\xi \chi }} &=&k^{\mu \nu \gamma |\alpha \beta \sigma ;\rho
\delta |\xi \chi },\qquad \frac{\partial \bar{W}^{\lambda \mu \nu \gamma
|\alpha \sigma }}{\partial r_{\rho \delta |\xi \chi }}=\bar{k}^{\lambda \mu
\nu \gamma |\alpha \sigma ;\rho \delta |\xi \chi },  \label{rt103} \\
\frac{\partial \bar{U}^{\mu \nu \gamma |\alpha \beta \sigma }}{\partial
t_{\rho \delta \xi |\chi }} &=&\hat{k}^{\mu \nu \gamma |\alpha \beta \sigma
;\rho \delta \xi |\chi },\qquad \frac{\partial \bar{W}^{\lambda \mu \nu
\gamma |\alpha \sigma }}{\partial t_{\rho \delta \xi |\chi }}=\tilde{k}%
^{\lambda \mu \nu \gamma |\alpha \sigma ;\rho \delta \xi |\chi },
\label{rt104}
\end{eqnarray}%
where $k^{\mu \nu \gamma |\alpha \beta \sigma ;\rho \delta |\xi \chi }$,\ $%
\bar{k}^{\lambda \mu \nu \gamma |\alpha \sigma ;\rho \delta |\xi \chi }$,\ $%
\hat{k}^{\mu \nu \gamma |\alpha \beta \sigma ;\rho \delta \xi |\chi }$, and $%
\tilde{k}^{\lambda \mu \nu \gamma |\alpha \sigma ;\rho \delta \xi |\chi }$
denote some non-derivative, constant tensors. By means of relations (\ref%
{rt103}) and (\ref{rt104}) we obtain (up to some irrelevant constants)%
\begin{eqnarray}
\bar{U}^{\mu \nu \gamma |\alpha \beta \sigma } &=&k^{\mu \nu \gamma |\alpha
\beta \sigma ;\rho \delta |\xi \chi }r_{\rho \delta |\xi \chi }+\hat{k}^{\mu
\nu \gamma |\alpha \beta \sigma ;\rho \delta \xi |\chi }t_{\rho \delta \xi
|\chi },  \label{rt105} \\
\bar{W}^{\lambda \mu \nu \gamma |\alpha \sigma } &=&\bar{k}^{\lambda \mu \nu
\gamma |\alpha \sigma ;\rho \delta |\xi \chi }r_{\rho \delta |\xi \chi }+%
\tilde{k}^{\lambda \mu \nu \gamma |\alpha \sigma ;\rho \delta \xi |\chi
}t_{\rho \delta \xi |\chi }.  \label{rt106}
\end{eqnarray}%
From the expression of $\omega _{2}$ given by (\ref{tww81}) we notice that
the terms $k^{\mu \nu \gamma |\alpha \beta \sigma ;\rho \delta |\xi \chi
}r_{\rho \delta |\xi \chi }$ and $\tilde{k}^{\lambda \mu \nu \gamma |\alpha
\sigma ;\rho \delta \xi |\chi }t_{\rho \delta \xi |\chi }$ appearing in (\ref%
{rt105}) and (\ref{rt106}) bring no contributions to cross-interactions. For
this reason, we take%
\begin{equation}
k^{\mu \nu \gamma |\alpha \beta \sigma ;\rho \delta |\xi \chi }=0,\qquad
\tilde{k}^{\lambda \mu \nu \gamma |\alpha \sigma ;\rho \delta \xi |\chi }=0,
\label{rt107}
\end{equation}%
such that (up to a total, irrelevant divergence) $\omega _{2}$ takes the form%
\begin{equation}
\omega _{2}=\bar{k}^{\lambda \mu \nu \gamma |\alpha \sigma ;\rho \delta |\xi
\chi }K_{\lambda \mu \nu \gamma |\alpha \sigma }r_{\rho \delta |\xi \chi }+%
\hat{k}^{\mu \nu \gamma |\alpha \beta \sigma ;\rho \delta \xi |\chi }F_{\mu
\nu \gamma |\alpha \beta \sigma }t_{\rho \delta \xi |\chi }.  \label{rt108}
\end{equation}%
The most general expression of $\bar{k}^{\lambda \mu \nu \gamma |\alpha
\sigma ;\rho \delta |\xi \chi }$ is represented by%
\begin{eqnarray}
\bar{k}^{\lambda \mu \nu \gamma |\alpha \sigma ;\rho \delta |\xi \chi }
&=&\kappa \left[ \frac{1}{4}\varepsilon ^{\lambda \mu \nu \gamma \rho \delta
}\left( \sigma ^{\xi \alpha }\sigma ^{\chi \sigma }-\sigma ^{\xi \sigma
}\sigma ^{\chi \alpha }\right) \right.   \notag \\
&&+\frac{1}{4}\varepsilon ^{\lambda \mu \nu \gamma \xi \chi }\left( \sigma
^{\rho \alpha }\sigma ^{\delta \sigma }-\sigma ^{\rho \sigma }\sigma
^{\delta \alpha }\right)   \notag \\
&&\left. -\frac{1}{24}\varepsilon ^{\lambda \mu \nu \gamma \left[ \rho
\delta \right. }\delta _{\ \tau }^{\xi }\delta _{\ \theta }^{\left. \chi %
\right] }\left( \sigma ^{\tau \alpha }\sigma ^{\theta \sigma }-\sigma ^{\tau
\sigma }\sigma ^{\theta \alpha }\right) \right] ,  \label{rt109}
\end{eqnarray}%
which then yields
\begin{equation}
\bar{k}^{\lambda \mu \nu \gamma |\alpha \sigma ;\rho \delta |\xi \chi
}K_{\lambda \mu \nu \gamma |\alpha \sigma }r_{\rho \delta |\xi \chi }=\kappa
\varepsilon ^{\lambda \mu \nu \gamma \rho \delta }r_{\rho \delta |\xi \chi
}K_{\lambda \mu \nu \gamma |}^{\qquad \xi \chi },  \label{rt110}
\end{equation}%
with $\kappa $ a real constant. On the other hand, there exist non-trivial
constant tensors of the type $\hat{k}^{\mu \nu \gamma |\alpha \beta \sigma
;\rho \delta \xi |\chi }$, but they all lead in the end to%
\begin{equation}
\hat{k}^{\mu \nu \gamma |\alpha \beta \sigma ;\rho \delta \xi |\chi }F_{\mu
\nu \gamma |\alpha \beta \sigma }t_{\rho \delta \xi |\chi }\equiv 0
\label{rt111}
\end{equation}%
due to the algebraic Bianchi I identity $F_{[\mu \nu \gamma |\alpha ]\beta
\sigma }=0$. Such constants have an intricate and non-illuminating form, and
therefore we will skip them. Inserting (\ref{rt110}) and (\ref{rt111}) in (%
\ref{rt108}), we deduce%
\begin{equation}
\omega _{2}=\kappa \varepsilon ^{\lambda \mu \nu \gamma \rho \delta }r_{\rho
\delta |\xi \chi }K_{\lambda \mu \nu \gamma |}^{\qquad \xi \chi }.
\label{rt112}
\end{equation}%
Acting with $\gamma $ on (\ref{rt112}), it is easy to see that%
\begin{equation}
\gamma \omega _{2}=-2\kappa \varepsilon ^{\lambda \mu \nu \gamma \rho \delta
}\left( \partial _{\left[ \lambda \right. }K_{\mu \nu \gamma ]|}^{\qquad \xi
}\right) \mathcal{C}_{\rho \delta |\xi }\neq \partial _{\mu }j_{2}^{\mu },
\label{rt113}
\end{equation}%
where $K_{\mu \nu \gamma |\tau }$ is the trace of the curvature tensor $%
K_{\mu \nu \gamma \alpha |\tau \beta }$, $K_{\mu \nu \gamma |\tau }=\sigma
^{\alpha \beta }K_{\mu \nu \gamma \alpha |\tau \beta }$. It is worthy to
notice that $\gamma \omega _{2}\neq \partial _{\mu }j_{2}^{\mu }$ follows
from the differential Bianchi II identity $\partial _{\beta }K_{\lambda \mu
\nu \gamma |}^{\qquad \beta \xi }=\partial _{\left[ \lambda \right. }K_{\mu
\nu \gamma ]|}^{\qquad \xi }$. Due to (\ref{rt113}), we must take%
\begin{equation}
\kappa =0,  \label{rt114}
\end{equation}%
and hence%
\begin{equation}
\omega _{2}=0.  \label{omega2}
\end{equation}%
Replacing (\ref{omega0}), (\ref{omega1}), and (\ref{omega2}) in (\ref{tww60}%
), we finally have that
\begin{equation}
\bar{a}_{0}^{\mathrm{t-r}}=0  \label{tv102}
\end{equation}%
in (\ref{rt73}).

Putting together the results expressed by formulae (\ref{rtn}), (\ref{rt43}%
), (\ref{rt55}), (\ref{rt69})--(\ref{rt70}), (\ref{rt73}), and (\ref{tv102}%
), we can state that the most general form of the first-order deformation
associated with the free theory (\ref{rt1b}) reads as%
\begin{eqnarray}
S_{1} &=&\int d^{6}x\left[ \varepsilon _{\mu \nu \alpha \lambda \beta \gamma
}\eta ^{\ast \mu \nu \alpha }\partial ^{\lambda }\mathcal{C}^{\beta \gamma
}\right.  \notag \\
&&-2\varepsilon _{\lambda \mu \nu \rho \beta \gamma }t^{\ast \lambda \mu \nu
|\alpha }\left( \partial ^{\rho }\mathcal{C}_{\quad \ \alpha }^{\beta \gamma
|}-\frac{1}{4}\delta _{\ \alpha }^{\gamma }\partial ^{\left[ \rho \right. }%
\mathcal{C}_{\quad \;\;\tau }^{\left. \beta \tau \right] |}\right)  \notag \\
&&\left. -2\varepsilon ^{\lambda \mu \nu \alpha \beta \gamma }t_{\lambda \mu
\nu |\rho }\left( \partial _{\sigma }\partial _{\alpha }r_{\beta \gamma
|}^{\quad \sigma \rho }-\frac{1}{2}\delta _{\ \gamma }^{\rho }\partial
^{\tau }\partial _{\alpha }r_{\beta \tau }\right) +r\right] .  \label{tv103}
\end{eqnarray}

\subsection{Higher-order deformations}

In the sequel we approach the higher-order deformation equations. The
second-order deformation is controlled by Eq. (\ref{tv65}). After some
computations we arrive at%
\begin{equation}
\left( S_{1},S_{1}\right) =s\int d^{6}x\left( 10r^{\lambda \rho |\left[
\alpha \beta ,\gamma \right] }r_{\lambda \rho |\left[ \alpha \beta ,\gamma %
\right] }-12r_{\lambda \rho |}^{\quad \left[ \alpha \beta ,\rho \right]
}r_{\quad \ \left[ \alpha \beta ,\sigma \right] }^{\lambda \sigma |}\right) ,
\label{rt116}
\end{equation}%
such that%
\begin{equation}
S_{2}=\int d^{6}x\left( -5r^{\lambda \rho |\left[ \alpha \beta ,\gamma %
\right] }r_{\lambda \rho |\left[ \alpha \beta ,\gamma \right] }+6r_{\lambda
\rho |}^{\quad \left[ \alpha \beta ,\rho \right] }r_{\quad \ \left[ \alpha
\beta ,\sigma \right] }^{\lambda \sigma |}\right) .  \label{tv104}
\end{equation}%
Using (\ref{tv103})--(\ref{tv104}) in (\ref{tv66}), we determine the
third-order deformation as%
\begin{equation}
S_{3}=0.  \label{rt115}
\end{equation}%
Under these conditions, it is easy to see that all the remaining
higher-order deformation equations are fulfilled with the choice%
\begin{equation}
S_{k}=0,\qquad k>3.  \label{rt117}
\end{equation}%
Inserting now relations (\ref{tv60}) (for $D=6$), (\ref{tv103}), (\ref{tv104}%
), (\ref{rt115}), and (\ref{rt117}) into (\ref{tv61}), we find nothing but
formula (\ref{rt30}), which then implies relations (\ref{rt32})--(\ref{rt37}%
). This ends the proof of the main result.

\section{Conclusion}

In this paper we have developed a cohomological approach to the problem of
constructing consistent interactions between a massless tensor gauge field
with the mixed symmetry $\left( 3,1\right) $ and a purely spin-two field
with the mixed symmetry $\left( 2,2\right) $. Under the general assumptions
of analyticity of the deformations in the coupling constant, locality,
(background) Lorentz invariance, Poincar\'{e} invariance, and the
requirement that the interaction vertices contain at most two spatiotemporal
derivatives of the fields, we have exhausted all the consistent, non-trivial
couplings. Our final result is rather surprising since it enables
non-trivial cross-couplings between the dual formulation of linearized
gravity in six spatiotemporal dimensions and the tensor field with the mixed
symmetry of the Riemann tensor. Although the cross-couplings break the PT
invariance and are merely mixing-component terms, still this is the first
situation encountered so far where the gauge transformations and
reducibility functions in the $\left( 3,1\right) $ sector are modified with
respect to the free ones.

\section*{Acknowledgments}

One of the authors (E.M.B.) acknowledges financial support from the contract
464/2009 in the framework of the programme IDEI of C.N.C.S.I.S. (Romanian
National Council for Academic Scientific Research).

\end{document}